\documentstyle[preprint,aps]{revtex}

\tightenlines 
\begin{document}

\draft  
\title{Predictive SUSY GUT Model for \\ CP Violation, Fermion Masses and Mixings  }
\author{ K.C. Chou$^{\dag}$ and Y.L. Wu$^{\ddag}$ \footnote{  Project supported in part by 
Outstanding Young Scientist Research Found of China   } }
\address{$^{\dag}$Chinese Academy of Sciences, Beijing 100864,  China  \\ 
 \vspace{0.3cm}
 $^{\ddag}$Institute of Theoretical Physics, \ Chinese Academy of Sciences\\ 
P.O. Box 2735, Beijing, 100080,  P.R. China }

\maketitle

\begin{abstract}
CP violation, fermion masses and mixing angles including that of 
neutrinos are studied in an SUSY SO(10)$\times \Delta (48)\times$ U(1) 
model. The nonabelian SU(3) discrete family symmetry $\Delta(48)$ 
associated with a simple scheme of U(1) charge assignment on various 
fields concerned in superpotential leads to unique Yukawa coupling 
matrices with zero textures. Thirteen parameters involving masses and 
mixing angles in the quark and charged lepton sector are successfully 
predicted by only four parameters. The masses and mixing angles for the 
neutrino sector could also be predicted by constructing an appropriate 
heavy Majorana neutrino mass matrix without involving new parameters. 
It is found that the atmospheric neutrino deficit, the mass limit put by hot 
dark matter and the LSND $\bar{\nu}_{\mu} \rightarrow \bar{\nu}_{e}$ events 
may simultaneously be explained, but solar neutrino puzzle can be solved only 
by introducing a sterile neutrino. An additional parameter is added to obtain 
the mass and mixing of the sterile  neutrino. The hadronic parameters $B_{K}$ 
and $f_{B}\sqrt{B}$ are extracted from the observed $K^{0}$-$\bar{K}^{0}$ and 
$B^{0}$-$\bar{B}^{0}$  mixings respectively. The direct CP violation 
($\varepsilon'/\varepsilon$) in kaon decays and the three angles $\alpha$, 
$\beta$ and $\gamma$ of the unitarity triangle in the CKM matrix are also 
presented. More precise measurements of $\alpha_{s}(M_{Z})$, $|V_{cb}|$, 
$|V_{ub}/V_{cb}|$, $m_{t}$, as well as various CP violation and neutrino 
oscillation experiments will provide an important test for the present model 
and guide us to a more fundamental theory.
\end{abstract}

\newpage


\section{Introduction}
 
The standard model (SM) is a great success. But 18 phenomenological 
parameters in the SM have to be introduced to describe all the low energy data
in the quark and charged lepton sector, which implies that the SM cannot be 
the fundamental theory. Furthermore, neutrinos are assumed to be massless in the SM.
Thus recent evidence for oscillation of atmospheric
neutrinos (and hence nonzero neutrino mass) reported by the Super-Kamiokande 
collaboration is thought as a major milestone in the search for new physics beyond 
the standard model(SM). Studies on neutrino physics have resulted in the following 
 observations: i), The Super-Kamiokande data\cite{SUPERK1} on 
 atmospheric neutrino anomaly provide a strong evidence that 
 neutrinos are massive; ii), The Super-Kamiokande data on 
 solar neutrino\cite{SUPERK2} cannot yet decisively establish 
  whether the solar neutrino deficit results from MSW solutions\cite{MSW} with 
  small or large mixing angles or Vacuum 
  oscillation solution though the most recent data
favor the MSW solutions with large mixing angle \cite{SUPERK3}. iii), To describe 
  all the neutrino phenomena such as the atmospheric neutrino anomaly, 
  the solar neutrino deficit and the results from the LSND experiment, 
  it is necessary to introduce a sterile neutrino. 
  It indicates that with only three light neutrinos, one of the 
  experimental data must be modified; iv), The current experimental data 
  cannot establish whether neutrinos are Dirac-type or Majorana-type. 
  The failure of detecting neutrinoless double beta decay only provides, 
  for Majorana-type neutrinos, an upper bound on an `effective' 
  electron neutrino mass; v), Large neutrino masses around several 
  electron volts may play an important role in the evolution of 
  the large-scale structure of the universe. On the other hand,
the 18 parameters in SM have  been extracted from various experiments 
although they are not yet
equally well known. Some of them have an accurcy of better than $1\%$, 
but some others less than $10 \%$. To improve the accuracy of
these parameters and understand them is a big challenge for 
particle physics. The mass spectrum and the mixing angles observed 
remind us that we are in a stage similar to that of atomic spectroscopy before 
Balmer. Much effort has been made along this direction. It 
was first observed by Gatto {\it et al}, 
Cabbibo and Maiani\cite{CM} that the Cabbibo angle is close to $\sqrt{
m_{d}/m_{s}}$. This observation initiated the investigation of the 
texture structure with zero elements \cite{ZERO} 
in the fermion Yukawa coupling matrices. The well-known examples are the
Fritzsch ansatz\cite{FRITZSCH} and Georgi-Jarlskog texture\cite{GJ}, 
which has been extensively studied and improved substantially in the 
literature\cite{TEXTURE}.  Ramond, Robert and Ross \cite{RRR} presented 
a general analysis on five symmetric texture structures with zeros in
the quark Yukawa coupling matrices. A general analysis and review of the 
previous studies on the texture structure was given by Raby 
in \cite{RABY}.  A numerous of papers\cite{SUSYSO} have investigated 
some interesting models with texture zeros based on supersymmetric 
(SUSY) SO(10).  A general operator analysis for the quark and charged lepton Yukawa coupling 
matrices with two zero textures `11' and `13' was made in ref. \cite{OPERATOR}. 
Though the texture `22' and `32' are not unique they
could fit successfully the 13 observables in the quark and charged lepton
sector with only six parameters.
We have also shown\cite{CHOUWU} that  the same
13 parameters as well as 10 parameters concerning 
the neutrino sector (though not unique for this sector) can be successfully 
described in an SUSY SO(10)$\times \Delta(48)\times$ U(1) model
with large $\tan\beta$, where the universality of Yukawa coupling of 
superpotential was assumed.  The resulting texture of mass matrices 
in the low energy region is quite unique and depends only on a 
single coupling constant and some vacuum expectation values (VEVs) 
caused by necessary symmetry breaking. The 23 parameters were 
predicted by only five parameters with three of them determined  
by the symmetry breaking scales of U(1), SO(10), SU(5) and SU(2)$_{L}$.
In that model, the ratio of the VEVs of two light Higgs $\tan\beta \equiv
v_{2}/v_{1}$ has large value  $\tan \beta \sim m_{t}/m_{b}$.  
 In general, there exists another 
interesting solution with small value of $\tan \beta \sim 1 $. Such a class of  
model could also give  a consistent prediction on top quark  mass and 
other low energy parameters. Furthermore, models with small value of 
$\tan \beta \sim 1 $ are of phenomenological interest in testing Higgs 
sector in the minimum supersymmetric standard model (MSSM)
at the  Colliders\cite{ELLIS}. Most of the 
existing models with small values of $\tan \beta$ in 
the literature have more parameters than those with 
large values of $\tan \beta \sim m_{t}/m_{b}$.  This
is because the third family unification condition 
$\lambda_{t}^{G} = \lambda_{b}^{G} = \lambda_{\tau}^{G}$ has been changed to 
 $\lambda_{t}^{G} \neq \lambda_{b}^{G} = \lambda_{\tau}^{G}$. Besides, 
some relations between the up-type and down-type quark (or charged lepton) mass 
matrices have also been lost in the small $\tan \beta$ case when two 
light Higgs doublets needed for SU(2)$_{L}$ symmetry breaking belong 
to different 10s of SO(10).  Although models with large $\tan \beta$ 
have less parameters,  large radiative corrections \cite{THRESHOLD} 
to the bottom quark mass and Cabibbo-Kobayashi-Maskawa (CKM) 
mixing angles might arise depending on 
an unkown spectrum of supersymmetric particles. 
  
   In a recent Rapid Communication\cite{CHOUWU2}, 
we have presented an alternative model with 
small value of  $\tan \beta \sim 1 $ based on
the same symmetry group SUSY SO(10)$\times \Delta(48)\times $U(1)
as the model \cite{CHOUWU} with large value of $\tan \beta $. It is
amazing to find  out that the model with small 
$\tan \beta \sim 1 $ in \cite{CHOUWU2} has more predictive power on fermion 
masses and mixings.  For convenience, we refer the model in \cite{CHOUWU} 
as Model I (with large $\tan \beta \sim m_{t}/m_{b}$) and the 
model in \cite{CHOUWU2,CHOUWU3} as Model II (with small $\tan \beta \sim 1 $).  

 In this paper, we will present in much greater detail an analysis for 
the model II. Our main considerations can be summarized as follows:

 1) SO(10) is chosen as the unification group\footnote{Recently, 
a three-family SO(10) grand unification theory was found  
in the string theories from orbifold approach \cite{GUST}. 
Other possible theories can be found from the free fermionic approach\cite{GUST2}.} 
so that the quarks and leptons
in each family are unified into a {\bf 16}-dimensional spinor representation 
of SO(10).

2) The non-abelian dihedral group $\Delta (48)$, 
a subgroup of SU(3) ($\Delta (3n^{2})$ with $n=4$), 
is taken as  the family group\footnote{Recently, the non-abelian discrete symmetry group of
a subgroup of U(2) has been taken as the family symmetry and applied to build the unified
 model\cite{ACL}}. Thus, the three families can be unified into 
a triplet {\bf 16}-dimensional spinor representation 
of SO(10)$\times \Delta (48)$.
U(1) is family-independent and is introduced to distinguish 
various fields which belong to the same representations of 
SO(10)$\times \Delta (48)$. The irreducible representations of 
$\Delta (48)$ consisting of five triplets and
three singlets are found to be sufficient to build  interesting texture
structures for fermion mass matrices. The symmetry $\Delta (48) \times$
U(1) naturally ensures the texture structure with zeros for
fermion Yukawa coupling matrices. Furthermore, the non-abelian flavor 
symmetries provides a super-GIM mechanism to supress  
flavor changing neutral currents induced by supersymmetric 
particles \cite{LNS,DLK,KS,FK}.
  
3) The universality of Yukawa coupling of the
superpotential before symmetry breaking is simply  assumed  to reduce  
possible free parameters, i.e., all the coupling coefficients in the
renormalizable superpotentials are assumed to be
equal and have the same origins from perhaps a more fundamental theory. We know
in general that universality of charges occurs only in 
the gauge interactions due to charge conservation 
like the electric charge of different particles. In the absence
of strong interactions, family symmetry could keep the universality of weak
interactions in a good approximation after breaking. 
In the present theory, there are very rich structures above 
the grand unification theory (GUT) 
scale with many heavy fermions and scalars and their interactions are taken to
be universal before symmetry breaking. 
All heavy fields must have some reasons to exist and interact which 
we do not understand at this moment. So that it can only be an ansatz at the
present moment since we do not know the answer governing the behavior of
nature above the GUT scale. As the Yukawa coupling matrices of the quarks and
leptons in the present model are generated at the GUT scale, so that the 
initial conditions of the renormalization group evaluation for them will be set 
at the GUT scale\footnote{For models in which the third family Yukawa
interaction is considered to be a renormalizble one starting from the Planck scale 
and the other two family Yukawa interactions are effectively generated at the GUT 
scale, one then needs to consider the renormalization effect the third family Yukawa 
coupling from the Planck scale down to the GUT
scale.}.  As the resulting Yukawa 
couplings only rely on the ratios of the coupling constants of the 
renormalizable superpotentials at the GUT scale, the 
predictions on the low energy observables will not be affected by the 
renormalization group (RG) effects running from the Planck scale to the GUT 
scale as long as the relative value of the ratios for 
the `22' and `32' textures is unchanged. For this aim, 
the `22' and `32' textures are constructed in such a way that they  
have a similar superpotential structure and 
the fields concerned belong to the same 
representations of the symmetry group. As we know that the renormalization 
group evaluation does not change the represenations of a symmetry group, 
thus the ratios of the coupling constants for the `22' and `32' textures 
should remain equal at the GUT scale. As we will see below, 
even if we abondon the general assumption of an unversal coupling for all the
Yukawa terms in the superpotential, the above feature can still be ensured 
by imposing a permutation symmetry among the fields concerning 
the `22' and `32' textures after family symmetry breaking.
As the numerical predictions on the low energy parameters so found are 
very encouraging and interesting, we believe that there must be a deeper 
reason that has to be found in the future.
 
 4)  The two light Higgs doublets are assumed to belong 
to an unique 10 representation Higgs of  SO(10).
 
 5) Both the symmetry breaking direction of SO(10) down to SU(5) 
and the two symmetry breaking directions of SU(5) down to SU(3)$_c \times$
SU(2)$_L \times$ U(1) are carefully chosen to ensure 
the needed Clebsch coefficients for quark and lepton mass matrices . 
The mass splitting between the up-type quark and down-type quark 
(or charged lepton) Yukawa 
couplings is attributed to the Clebsch factors caused by the SO(10) symmetry 
breaking direction. Thus the third family four-Yukawa coupling relation 
at the GUT scale will be given by
\begin{equation}
\lambda_{b}^{G} = \lambda_{\tau}^{G} = \frac{1}{3^n} \lambda_{t}^{G} =
 5^{n+1} \lambda_{\nu_{\tau}}^{G}
\end{equation}
where the factors $1/3^{n}$ and $5^{n+1}$ with $n$ being an integer are the
Clebsch factors. A factor $1/3^{n}$ will also
multiply the down-type quark and charged lepton Yukawa 
coupling matrices. 
 
6) CP symmetry is broken spontaneously in the model, 
a maximal CP violation is assumed to further diminish free parameters.
 
With the above considerations, 
the resulting model has found to provide a successful prediction 
on 13 parameters in 
the quark and charged lepton sector as well as an interesting prediction on 
10 parameters in the neutrino sector with only four parameters. 
One is the  universal coupling constant 
and the other three are determined by the ratios of 
vacuum expectation values (VEVs) 
of the symmetry breaking scales and the RG effects above the GUT scale.
One additional parameter resulting from the VEV of a singlet scalar is
introduced to obtain the mass and mixing angle of a sterile neutrino. 
Our paper is organized as follows: 
In section 2, we will present the 
results of the Yukawa coupling matrices. The resulting masses and CKM 
quark mixings are presented in section 3. In section 4 neutrino masses and 
CKM-type mixings in the lepton sector are presented. All existing 
neutrino experiments are discussed and shown how they may be understandable in the
present model.  Conclusions and remarks 
are presented in the last section. 

\section{Yukawa Coupling Matrices}

 With the above considerations, a model based on the symmetry 
group SUSY SO(10)$\times \Delta(48) \times$ U(1) with 
a single coupling constant and small 
value of $\tan \beta$  is constructed. 
 Yukawa coupling matrices which determine the masses and mixings
of all quarks and leptons are obtained by carefully choosing the 
structure of the physical vacuum and integrating out the heavy fermions at the
GUT scale. We find
\begin{equation}
\Gamma_{u}^{G} = \frac{2}{3}\lambda_{H}  \left(\begin{array}{ccc} 
0  &  \frac{3}{2}z'_{u} \epsilon_{P}^{2} &   0   \\
\frac{3}{2}z_{u} \epsilon_{P}^{2} &  - 3 y_{u} \epsilon_{G}^{2} e^{i\phi} 
 & -\frac{\sqrt{3}}{2}x_{u}\epsilon_{G}^{2}  \\
0  &  - \frac{\sqrt{3}}{2}x_{u}\epsilon_{G}^{2}  &  w_{u} 
\end{array} \right)
\end{equation}   
and
\begin{equation}
 \Gamma_{f}^{G} = \frac{2}{3}\lambda_{H} \frac{(-1)^{n+1}}{3^{n}} 
\left( \begin{array}{ccc} 
0  &  -\frac{3}{2}z'_{f} \epsilon_{P}^{2} &   0   \\
-\frac{3}{2}z_{f} \epsilon_{P}^{2} &  3 y_{f} 
\epsilon_{G}^{2} e^{i\phi}  
& -\frac{1}{2}x_{f}\epsilon_{G}^{2}  \\
0  &  -\frac{1}{2}x_{f}\epsilon_{G}^{2}  &  w_{f} 
\end{array} \right)
\end{equation}   
for $f=d,e$,  and 
\begin{equation}
\Gamma_{\nu}^{G} = \frac{2}{3}\lambda_{H}\frac{(-1)^{n+1}}{3^n}
\frac{1}{5^{n+1}} 
\left( \begin{array}{ccc} 
0  &  -\frac{15}{2}z'_{\nu} \epsilon_{P}^{2} &   0   \\
-\frac{15}{2}z_{\nu} \epsilon_{P}^{2} &  15 y_{\nu} 
\epsilon_{G}^{2} e^{i\phi}  
& -\frac{1}{2}x_{\nu}\epsilon_{G}^{2}  \\
0  &  -\frac{1}{2}x_{\nu}\epsilon_{G}^{2}  &  w_{\nu} 
\end{array} \right)
\end{equation}   
for Dirac-type neutrino coupling.  We will choose $n=4$ in the following 
considerations.
$\lambda_{H}=\lambda_{H}^{0}r_{3}$, $\epsilon_{G}\equiv (\frac{v_{5}}{v_{10}})
\sqrt{\frac{r_{2}}{r_{3}}}$ and $\epsilon_{P}\equiv
(\frac{v_{5}}{\bar{M}_{P}})\sqrt{\frac{r_{1}}{r_{3}}}$ are three parameters. 
Where $\lambda_{H}^{0}$ is a universal 
coupling constant expected to be of order one, $r_{1}$, $r_{2}$ and $r_{3}$ 
denote the ratios of the coupling constants of the superpotential at 
the GUT scale for the textures `12', `22' (`32') and `33' respectively. 
They represent the possible renormalization group (RG) effects 
running from the scale $\bar{M}_{P}$ to the GUT scale. Note that the RG effects
for the  textures `22' and `32' are considered to be the same since they are
generated from a similar superpotential structure after integrating out the
heavy fermions and the fields concerned belong to 
the same representations of the symmetry group. This can be explicitly seen 
from their effective operators $W_{22}$ and $W_{32}$ given in eq. (6). 
$\bar{M}_{P}$, $v_{10}$ and $v_{5}$ are the VEVs for 
U(1)$\times \Delta(48)$, SO(10) and SU(5) symmetry breaking
respectively. $\phi$ is the physical CP phase\footnote{ We have rotated
away other possible phases by a phase redefinition of the fermion fields.} 
arising from the VEVs. The assumption of maximum CP violation implies that 
$\phi = \pi/2$. $x_{f}$, $y_{f}$, $z_{f}$, and $w_{f}$ $(f = u, d, e, \nu)$ 
are the Clebsch factors of SO(10) determined by the 
directions of symmetry breaking of the adjoints {\bf 45}'s. 
The following three directions have been chosen for symmetry breaking, 
namely:
\begin{eqnarray}
& & <A_{X}>=2v_{10}\  diag. (1,\ 1,\ 1,\ 1,\ 1)\otimes \tau_{2}, \nonumber \\ 
& & <A_{z}> =2v_{5}\  diag. (-\frac{1}{3},\ -\frac{1}{3},\ -\frac{1}{3},\ 
-1,\ -1)\otimes \tau_{2}, \\
& & <A_{u}>=\frac{1}{\sqrt{3}}v_{5}\  
diag. (2,\ 2,\ 2,\ 1,\  1)\otimes \tau_{2}  \nonumber
\end{eqnarray} 
Their corresponding U(1) hypercharges are given in Table I.

{\bf TABLE I.}  U(1) Hypercharge Quantum Number 
\\

\begin{tabular}{|c|c|c|c|c|c|}  \hline  
   & `X' & `u'   &  `z'  &  B-L  &  $T_{3R}$  \\
$q$  & 1   &  $\frac{1}{3}$  &  $\frac{1}{3}$  & $\frac{1}{3}$  & 0  \\
$u^{c}$ &   1  &  0 & $\frac{5}{3}$  & -$\frac{1}{3}$ & $\frac{1}{2}$  \\
$d^{c}$  & -3  & -$\frac{2}{3}$  & -$\frac{7}{3}$  & -$\frac{1}{3}$ & 
-$\frac{1}{2}$   \\
$l$  & - 3 & -1 & -1 & -1 & 0  \\
$e^{c}$  & 1  & $\frac{2}{3}$  &  -1  & 1 &  -$\frac{1}{2}$ \\     
$\nu^{c}$  &  5  & $\frac{4}{3}$  &  3  & 1 & $\frac{1}{2}$ \\   \hline
\end{tabular}
\\ 

 The Clebsch factors associated with the symmetry breaking directions can be 
easily read off from the U(1) hypercharges of the above table. The
related effective operators obtained after the heavy 
fermion pairs integrated out are\footnote{Note that $W_{22}$ 
is slightly modified in comparison with the one in \cite{CHOUWU2} 
since we have renormalized the VEV 
$<A_{u}>$. As a consequence, only the Clebsch factor $y_{\nu}$ is modified,
which does not affect all the numerical predictions.}
\begin{eqnarray} 
W_{33} & = & \lambda_{H}^{0}r_{3}16_{3} \ \eta_{X}\eta_{A}10_{1}
\eta_{A}\eta_{X}\ 16_{3}  
\nonumber \\
W_{32} & = & \lambda_{H}^{0}r_{2}16_{3}\ \eta_{X}\eta_{A} 
\left(\frac{A_{z}}{A_{X}}\right)10_{1}\left(\frac{A_{z}}{A_{X}}\right) 
\eta_{A}\ 16_{2}  \nonumber \\
W_{22} & = &  \lambda_{H}^{0}r_{2}16_{2}\ \eta_{A}
\left(\frac{A_{u}}{A_{X}}\right) 10_{1}
\left(\frac{A_{u}}{A_{X}}\right)\eta_{A}\ 16_{2}e^{i\phi}  \\ 
W_{12} & = & \lambda_{H}^{0}r_{1}\  
16_{1}\  [ \left(\frac{v_{5}}{\bar{M}_{P}}\right)^{2}\eta'_{A}10_{1}\eta'_{A}  
\nonumber \\
 & & + \left(\frac{v_{10}}{\bar{M}_{P}}\right)^{2}\eta_{A}
\left(\frac{A_{u}}{A_{X}}\right)10_{1} \left(\frac{A_{z}}{A_{X}}\right) 
\eta_{A} ]\ 16_{2}    \nonumber  
\end{eqnarray}
with $n=4$ and $\phi = \pi/2$. $\eta_{A} = (v_{10}/A_{X})^{n+1}$ and 
$\eta'_{A} = (v_{10}/A_{X})^{n-3}$.
The factor $\eta_{X} = 1/\sqrt{1 + 2\eta_{A}^{2}}$  in eq.
(6) arises from mixing, and provides a factor of $1/\sqrt{3}$ for the up-type 
quark. It remains almost unity for the down-type quark and charged lepton
as well as neutrino due to the suppression of large 
Clebsch factors in the second
term of the square root. The relative phase (or sign) between  the
two terms in the operator $W_{12}$ has been fixed.
The resulting Clebsch  factors are
\begin{eqnarray}
& & w_{u}=w_{d}=w_{e}=w_{\nu} =1, \nonumber \\ 
& & x_{u}= 5/9,\  x_{d}= 7/27,\  x_{e}=-1/3,\  x_{\nu} = 1/5, \nonumber \\  
& & y_{u}=0, \  y_{d}=y_{e}/3=2/27, \ y_{\nu} = 4/225, \\ 
& & z_{u}=1, \  z_{d}=z_{e}= -27,\   z_{\nu} = -15^3 = -3375, \nonumber \\
& &  z'_u = 1-5/9 = 4/9,\  z'_d = z_d + 7/729 \simeq z_{d},\nonumber \\
& &   z'_{e} = z_{e} - 1/81 \simeq z_{e},\   
z'_{\nu} = z_{\nu} + 1/15^{3} \simeq z_{\nu}.  \nonumber  
\end{eqnarray}
In obtaining the $\Gamma_{f}^{G}$ matrices, 
some small terms  arising from mixings between the chiral 
fermion $16_{i}$ and the heavy fermion pairs $\psi_{j} (\bar{\psi}_{j})$ are 
neglected. They are  expected to change the numerical results no more than 
a few percent for the up-type quark mass matrix and are negligible for the 
down-type quark and lepton mass matrices due to the strong suppression of the
Clebsch factors. This set of effective operators which lead to 
the above given Yukawa coupling matrices $\Gamma_{f}^{G}$ is quite unique
for a successful prediction on fermion masses and mixings.
A general superpotential leading to the above effective operators 
will be given in  section 6.  We would like to point out that 
unlike many other models in which  $W_{33}$ is assumed to be 
a renormalizable interaction before symmetry breaking, the Yukawa couplings 
of all the quarks and leptons (both heavy and light) in both Model II and 
Model I are generated at the GUT scale after the breakdown of 
the family group and SO(10). Therefore, initial
conditions for renormalization group (RG) evolution  will be set at the
GUT scale for all the quark and lepton Yukawa 
couplings. The hierarchy among the three families is described 
by the two ratios $\epsilon_{G}$ and $\epsilon_{P}$. 
The mass splittings between the quarks and leptons as well as between the 
up and down quarks are determined by the Clebsch factors of SO(10).
From the GUT scale down to low energies, Renormalization Group (RG) 
evolution has been taken into account.
The top-bottom splitting in the present model is mainly 
attributed to the Clebsch factor $1/3^{n}$ with $n=4$ 
rather than the large value of $\tan\beta$ caused by the hierarchy of 
the VEVs $v_{1}$ and $v_{2}$ of the two light Higgs doublets.  
 
 An adjoint {\bf 45} $A_{X}$ and a 16-dimensional representation 
Higgs field $\Phi$ ($\bar{\Phi}$) are needed for breaking SO(10) down to SU(5). 
Adjoint {\bf 45} $A_{z}$ and $A_{u}$ are needed to break SU(5) further 
down to the standard model SU(3)$_{c} \times$ SU$_{L}(2) \times$ U(1)$_{Y}$.

\section{Predictions}

  From the Yukawa coupling matrices given above with $n=4$ and $\phi = \pi/2$,
the 13 parameters in the SM can be determined by only four parameters: a
universal coupling constant $\lambda_{H}$ and three ratios: 
$\epsilon_{G}$, $\epsilon_{P}$ and 
$\tan \beta = v_2/v_1 $. In obtaining physical masses and mixings, 
renormalization group (RG) effects below the GUT scale 
has been further taken into consideration.
The result  at the low energy obtained by scaling down from the GUT 
scale will depend on the strong coupling constant $\alpha_{s}$.   
From low-energy measurements\cite{LEM}  and 
lattice calculations\cite{ALPHAS}, 
$\alpha_{s}$ at the scale $M_{Z}$, has value around
$\alpha_{s}(M_{Z})=0.113$, which was also found to be consistent with a recent 
global fit \cite{FIT} to the LEP data.  This value might be 
reached in nonminimal SUSY GUT models through
large threshold effects. As our focus here is on the fermion masses and
mixings, we shall not discuss it  in this paper.  In the present 
consideration, we  take $\alpha_{s}(M_{Z})\simeq 0.113$. 
The prediction on fermion masses and mixings thus obtained is 
found to be remarkable.  Our numerical
predictions are given in Tables II and III with four 
input parameters, three of them are the well measured charged lepton massess and
another is the bottom quark mass.

 The  predictions on the quark masses 
and mixings as well as CP-violating effects presented in Table IIb agree 
remarkably with those extracted from various experimental data. Especially, 
there are four predictions on $|V_{us}|$, $|V_{ub}/V_{cb}|$, 
$|V_{td}/V_{ts}|$ and $m_{d}/m_{s}$  which are independent of  the
RG scaling (see  eqs. (41)-(44) below).

  Let us now analyze in detail the above predictions. 
To a good approximation, the up-type and down-type
quark Yukawa coupling matrices can be diagonalized in the form 
\begin{equation}
 V_d = \left( \begin{array}{ccc} 
c_1  &  s_1 &   0   \\
-s_1  &  c_1  & 0   \\
0  &  0 &  1
\end{array} \right) \left( \begin{array}{ccc} 
e^{-i\phi}  &  0  &   0   \\
0  &  c_d  & s_d   \\
0  &  -s_d &  c_d
\end{array} \right)
\end{equation}
\begin{equation}   
 V_u = \left( \begin{array}{ccc} 
c_2  &  s_2 &   0   \\
-s_2  &  c_2  & 0   \\
0  &  0 &  1
\end{array} \right) \left( \begin{array}{ccc} 
e^{-i\phi}  &  0  &   0   \\
0  &  c_u  & s_u   \\
0  &  -s_u &  c_u
\end{array} \right)
\end{equation}   
The CKM matrix at the GUT scale is then given by $V_{CKM} = V_u
V_{d}^{\dagger}$.  Where $s_{i} \equiv \sin(\theta_{i}) $ and $c_{i} \equiv 
\cos(\theta_{i})$ ($i=1,2,u,d$). For $\phi = \pi /2$, 
the angles $\theta_{i}$ at the GUT scale are given by
\begin{eqnarray} 
& & \tan(\theta_1) \simeq -\frac{z_{d}}{2y_{d}}
\frac{\epsilon_{P}^{2}}{\epsilon_{G}^{2}}, \qquad 
\tan(\theta_2) \simeq \frac{2w_{u}z_u}{x_{u}^{2}}
\frac{\epsilon_{P}^{2}}{\epsilon_{G}^{4}},    \\
& & \tan(\theta_d) \simeq \frac{x_{d}}{2w_{d}}\epsilon_{G}^{2}, 
\qquad  \tan(\theta_u) \simeq \frac{\sqrt{3}}{2}
\frac{x_u}{w_{u}}\epsilon_{G}^{2} .
\end{eqnarray} 
and the Yukawa eigenvalues at the GUT scale are found to be 
\begin{eqnarray} 
& & \frac{\lambda_{u}^{G}}{\lambda_{c}^{G}} = \frac{4w_{u}^{2}z_{u}z'_{u}}
{x_{u}^{4}\epsilon_{G}^{4}}
\frac{\epsilon_{P}^{4}}{\epsilon_{G}^{4}}, \qquad 
\frac{\lambda_{c}^{G}}{\lambda_{t}^{G}} = \frac{3}{4}\frac{x_{u}^{2}}
{w_{u}^{2}}\epsilon_{G}^{4},  \\
& & \frac{\lambda_{d}^{G}}{\lambda_{s}^{G}} 
\left(1- \frac{\lambda_{d}^{G}}{\lambda_{s}^{G}}\right)^{-2}  = 
\frac{z_{d}^{2}}{4y_{d}^{2}}
\frac{\epsilon_{P}^{4}}{\epsilon_{G}^{4}}, \qquad 
\frac{\lambda_{s}^{G}}{\lambda_{b}^{G}} = 3\frac{y_{d}}{w_{d}}
\epsilon_{G}^{2},  \\
& & \frac{\lambda_{e}^{G}}{\lambda_{\mu}^{G}} 
\left(1- \frac{\lambda_{e}^{G}}{\lambda_{\mu}^{G}}\right)^{-2}  = 
\frac{z_{e}^{2}}{4y_{e}^{2}}
\frac{\epsilon_{P}^{4}}{\epsilon_{G}^{4}}, \qquad 
\frac{\lambda_{\mu}^{G}}{\lambda_{\tau}^{G}} = 3\frac{y_{e}}{w_{e}}
\epsilon_{G}^{2}.
\end{eqnarray}
Using the eigenvalues and angles of these Yukawa matrices , 
one can easily find the following ten relations  among fermion masses 
and CKM matrix elements at the GUT scale
\begin{eqnarray}
& & \left(\frac{m_{b}}{m_{\tau}}\right)_{G} = 1,  \\
& & \left(\frac{m_{s}}{m_{\mu}}\right)_{G} = \frac{1}{3}, \qquad or \qquad
\left(\frac{m_{s}}{m_{b}}\right)_G = \frac{1}{3} 
\left(\frac{m_{\mu}}{m_{\tau}}\right)_{G}  \\
& & \left(\frac{m_{d}}{m_{s}}\right)_G 
\left(1- (\frac{m_{d}}{m_{s}})_G \right)^{-2} 
= 9 \left(\frac{m_{e}}{m_{\mu}}\right)_G
\left(1- \left(\frac{m_{e}}{m_{\mu}}\right)_G \right)^{-2} ,  \\
& & \left(\frac{m_{t}}{m_{\tau}}\right)_{G} = 81\  \tan \beta \ ,    \\
& & \left(\frac{m_{c}}{m_{t}}\right)_G = \frac{25}{48} 
  \left(\frac{m_{\mu}}{m_{\tau}}\right)_{G}^{2} \ ,  \\
& & \left(\frac{m_{u}}{m_{c}}\right)_G = \frac{4}{9} 
\left(\frac{4}{15}\right)^{4} 
  \left(\frac{m_{e}m_{\tau}^{2}}{m_{\mu}^{3}} \right)_G \ ,   \\
& & |\frac{V_{ub}}{V_{cb}}|_{G} = \tan (\theta_{2}) = 
\left(\frac{4}{15}\right)^{2} \left(\frac{m_{\tau}}{
m_{\mu}}\right)_G \sqrt{\left(\frac{m_{e}}{m_{\mu}}\right)_G} \ , \\
& & |\frac{V_{td}}{V_{ts}}|_{G} = \tan (\theta_{1}) = 
3 \sqrt{\left(\frac{m_{e}}{m_{\mu}}\right)_G} \ , \\
& & |{V_{us}}|_{G} = c_{1}c_{2} \sqrt{\tan^{2}(\theta_{1}) + \tan^{2}
(\theta_{2}) } =  3 \sqrt{\left(\frac{m_{e}}{m_{\mu}}\right)_G} 
\left( \frac{1 + (\frac{16}{675}(\frac{m_{\tau}}{m_{\mu}})_G)^{2}}{1 + 9
(\frac{m_{e}}{m_{\mu}})_G}\right)^{1/2} \\
& & |{V_{cb}}|_{G} = c_{2} c_{d}c_{u} (\tan (\theta_{u}) - \tan (\theta_{d}) )
= \frac{15\sqrt{3}- 7}{15\sqrt{3}}\frac{5}{4\sqrt{3}}
 \left(\frac{m_{\mu}}{m_{\tau}}\right)_G \  .
\end{eqnarray}
The Clebsch factors in eq. (7) appeared as those miraculus numbers 
in  the above relations.
The index `G' refers throughout to quantities  at the GUT scale.  
The first two relations are well-known in the Georgi-Jarlskog texture. 
The physical fermion masses and mixing angles are related to the above 
Yukawa eigenvalues and angles through 
 the renormalization group (RG) equations\cite{RG}. 
As most Yukawa couplings in the present model are much smaller than the top 
quark Yukawa coupling $\lambda_{t}^{G} \sim 1 $. In a good approximation, we
will only keep top quark Yukawa coupling terms in the RG equations and 
neglect all other Yukawa coupling terms in the RG equations.
 The RG evolution  will be described by
three kinds of  scaling factors. Two  of them ($\eta_{F}$  
and $R_{t}$ ) arise from running the Yukawa parameters from the GUT scale 
down to the SUSY breaking scale $M_{S}$ which is chosen to be close
to the top quark mass, i.e., $M_{S} \simeq m_{t}\simeq 170$ GeV, and 
are defined as
\begin{eqnarray}
& & m_{t} (M_{S}) = 
\eta_{U}(M_{S})\  \lambda_{t}^{G}\ R_{t}^{-6}\  
\frac{v}{\sqrt{2}} \sin \beta \ , \\
& &  m_{b} (M_{S}) = \eta_{D}(M_{S})\  
\lambda_{b}^{G}\ R_{t}^{-1}\  \frac{v}{\sqrt{2}} \cos \beta \  ,  \\
& & m_{i} (M_{S}) = \eta_{U} (M_{S})\  \lambda_{i}^{G}\ R_{t}^{-3} \   
\frac{v}{\sqrt{2}} \sin \beta \  , \qquad i=u, c,  \\
& & m_{i} (M_{S}) = \eta_{D}(M_{S}) \lambda_{i}^{G} 
\frac{v}{\sqrt{2}} \cos \beta \  , \qquad 
i = d, s,  \\
& & m_{i} (M_{S}) = \eta_{E}(M_{S}) \lambda_{i}^{G} 
\frac{v}{\sqrt{2}} \cos \beta \  , \qquad i=e, \mu , \tau , \\
& & \lambda_{i}(M_{S}) = \eta_{N}(M_S)\  \lambda_{i}^{G}\  R_{t}^{-3} \ ,  
\qquad i = \nu_{e}, \nu_{\mu}, \nu_{\tau} \  .
\end{eqnarray}
with $v=246$ GeV. $\eta_{F}(M_{S})$ and $R_{t}$ are given by 
\begin{eqnarray}
& & \eta_{F}(M_{S}) = \prod_{i=1}^{3}
\left(\frac{\alpha_{i}(M_{G})}{\alpha_{i}(M_{S})}\right)^{c_{i}^{F}/2b_{i}} , 
\qquad F=U, D, E, N  \\
& & R_{t}^{-1} = e^{-\int_{\ln M_{S}}^{\ln M_{G}} 
(\frac{\lambda_{t}(t)}{4\pi})^{2} dt } 
=(1 + (\lambda_{t}^{G})^{2} K_{t})^{-1/12} =
 \left(1- \frac{\lambda_{t}^{2}(M_{S})}{\lambda_{f}^{2}} \right)^{1/12}   
\end{eqnarray}
with $c_{i}^{U} = (\frac{13}{15}, 3, \frac{16}{3})$, 
$c_{i}^{D} = (\frac{7}{15}, 3, \frac{16}{3})$ ,  
$c_{i}^{E} = (\frac{27}{15}, 3, 0)$, $c_{i}^{N} = (\frac{9}{25}, 3, 0)$,  
and $b_{i} = (\frac{33}{5}, 1, -3)$,  
where $\lambda_{f}$ is the fixed  point value of $\lambda_{t}$ and is given by 
\begin{equation}
\lambda_{f} = \frac{2\pi \eta_{U}^{2} } {\sqrt{3I(M_{S})}}, 
\qquad I(M_{S}) = \int_{\ln M_{S}}^{\ln M_{G}} \eta_{U}^{2}(t) dt 
\end{equation} 
The factor $K_{t}$ is related to the fixed point value via $K_{t} =
\eta_{U}^{2} /\lambda_{f}^{2} = \frac{3 I(M_{S})}{4\pi^{2}}$ .  The numerical 
value for $I$ taken from Ref. \cite{FIX} is 113.8 
for $M_{S} \simeq m_{t} = 170$GeV. 
$\lambda_{f}$ cannot be equal to
$\lambda_{t}(M_{S})$ exactly , since that would correspond to infinite 
$\lambda_{t}^{G}$ , and lead to the 
so called Landau pole problem at the GUT scale.  
Other RG scaling factors are derived by
running Yukawa couplings below $M_{S}$
\begin{eqnarray}  
& & m_{i}(m_{i}) = \eta_{i} \  m_{i} (M_{S}), \qquad i = c,b , \\
& & m_{i}(1GeV) = \eta_{i}\  m_{i} (M_{S}), \qquad i = u,d,s
\end{eqnarray}
where $\eta_{i}$ are the renormalization factors. 
The physical top quark mass is given by 
\begin{equation} 
M_{t} = m_{t}(m_{t}) \left(1 + \frac{4}{3}\frac{\alpha_{s}(m_{t})}{\pi}\right)
\end{equation}
In numerical calculations, we take $\alpha^{-1}(M_Z) = 127.9 $, $s^{2}(M_Z) 
= 0.2319$, $M_Z = 91.187$ GeV and use the gauge 
couplings at $M_{G} \sim 2 \times 10^{16}$ GeV at GUT scale and that of 
 $\alpha_1$ and $\alpha_2$ at $M_S \simeq m_{t} \simeq 170$ GeV  
\begin{eqnarray}
& & \alpha_{1}^{-1}(m_t) = \alpha_{1}^{-1}(M_Z) + \frac{53}{30\pi} \ln
\frac{M_Z}{m_t} = 58.59 , \\
& & \alpha_{2}^{-1}(m_t) = \alpha_{2}^{-1}(M_Z) - \frac{11}{6\pi} \ln
\frac{M_Z}{m_t} = 30.02 , \\
& & \alpha_{1}^{-1}(M_G) = \alpha_{2}^{-1}(M_G) = \alpha_{3}^{-1}(M_G)
 \simeq 24 
\end{eqnarray}
we  keep $\alpha_{3}(M_Z)$ as a free parameter in this note.
The precise prediction on $\alpha_{3}(M_{Z})$ concerns 
GUT and SUSY threshold corrections. We shall not discuss 
it here since 
our focus in this note is the fermion masses and mixings. 
Including the three-loop QCD and one-loop QED contributions, 
the values of $\eta_{i}$ in Table IV  
will be used in  numerical calculations.

 It is  interesting to note that the mass ratios of the charged leptons are
almost independent of the RG scaling factors since $\eta_{e} = \eta_{\mu} 
= \eta_{\tau}$ (up to an accuracy $O(10^{-3})$), namely 
\begin{equation} 
\frac{m_{e}}{m_{\mu}} = \left(\frac{m_{e}}{m_{\mu}}\right)_{G}, \qquad 
\frac{m_{\mu}}{m_{\tau}} = \left(\frac{m_{\mu}}{m_{\tau}}\right)_{G}
 \end{equation}
which is different from the models with large $\tan \beta$. 
In the present model the $\tau$ lepton Yukawa coupling is small.   
It is  easily seen that four 
relations represented by eqs. (21)-(23) and (17) hold at 
low energies. Using the known lepton masses 
$m_{e}= 0.511$ MeV, $m_{\mu} = 105.66$ MeV, and 
$m_{\tau} = 1.777$ GeV, we obtain four important RG scaling-independent
predictions:
\begin{eqnarray}
& & |V_{us}| = |{V_{us}}|_{G} = \lambda 
\simeq  3 \sqrt{\frac{m_{e}}{m_{\mu}}} \left( \frac{1 +
(\frac{16}{675}\frac{m_{\tau}}{m_{\mu}})^{2}}{1 + 9
\frac{m_{e}}{m_{\mu}}}\right)^{1/2} = 0.22 , \\
& & |\frac{V_{ub}}{V_{cb}}|= |\frac{V_{ub}}{V_{cb}}|_{G} = \lambda
\sqrt{\rho^{2} + \eta^{2} } \simeq 
\left(\frac{4}{15}\right)^{2} \frac{m_{\tau}}{
m_{\mu}}\sqrt{\frac{m_{e}}{m_{\mu}}} = 0.083 , \\
& & |\frac{V_{td}}{V_{ts}}|= |\frac{V_{td}}{V_{ts}}|_{G} = \lambda 
\sqrt{ (1-\rho)^{2} + \eta^{2} } \simeq
3 \sqrt{\frac{m_{e}}{m_{\mu}}} = 0.209 , \\
& & \frac{m_{d}}{m_{s}} \left(1- \frac{m_{d}}{m_{s}} \right)^{-2} 
= 9 \frac{m_{e}}{m_{\mu}}
\left(1- \frac{m_{e}}{m_{\mu}} \right)^{-2} , \quad i.e.,  
\quad \frac{m_{d}}{m_{s}} = 0.040 
\end{eqnarray}
and six RG scaling-dependent predictions:
\begin{eqnarray}
& & |V_{cb}| = |{V_{cb}}|_{G} R_{t} = A\lambda^{2}
= \frac{15\sqrt{3}- 7}{15\sqrt{3}}\frac{5}{4\sqrt{3}}
 \frac{m_{\mu}}{m_{\tau}} R_{t}  = 0.0391 \ 
\left(\frac{0.80}{R_{t}^{-1}}\right), \\
& & m_{s}(1GeV) = \frac{1}{3} m_{\mu} \frac{\eta_{s}}{\eta_{\mu}} \eta_{D/E}
= 159.53\  \left(\frac{\eta_{s}}{2.2}\right) \left(\frac{\eta_{D/E}}{2.1} \right)\  MeV, \\
& & m_{b}(m_{b}) = m_{\tau} \frac{\eta_{b}}{\eta_{\tau}} \eta_{D/E} R_{t}^{-1} 
= 4.25 \  \left( \frac{\eta_{b}}{1.49}\right) 
\left( \frac{\eta_{D/E}}{2.04}\right) 
\left( \frac{R_{t}^{-1}}{0.80} \right) \  GeV \ , \\
& & m_{u}(1GeV) = \frac{5}{3}(\frac{4}{45})^{3} \frac{m_{e}}{m_{\mu}} \eta_{u} 
R_{t}^{3} m_{t} = 4.23\ \left(\frac{\eta_{u}}{2.2}\right)
\left(\frac{0.80}{R_{t}^{-1}}\right)^{3} 
\left( \frac{m_{t}(m_{t})}{174 GeV}\right)\  MeV \ , \\ 
& & m_{c}(m_{c}) = \frac{25}{48} (\frac{m_{\mu}}{m_{\tau}})^{2} 
\eta_{c} R_{t}^{3} m_{t} = 1.25\  \left(\frac{\eta_{c}}{2.0}\right)
\left(\frac{0.80}{R_{t}^{-1}}\right)^{3} 
\left( \frac{m_{t}(m_{t})}{174 GeV}\right)\ GeV ,  \\
& & m_{t}(m_{t}) = \frac{\eta_{U}}{\sqrt{K_{t}}} \sqrt{1 - R_{t}^{-12}}
\frac{v}{\sqrt{2}} \sin \beta  \nonumber \\
& & \qquad = 174.9\ \left( \frac{\sin \beta}{0.92}
\right)  \left( \frac{\eta_{U}}{3.33}\right) 
\left( \sqrt{ \frac{8.65}{K_{t}}}\right)  \left(
\frac{\sqrt{1-R_{t}^{-12}}}{0.965} \right) \  GeV   
\end{eqnarray}
We have used the fixed point property for the top quark mass. 
These predictions  depend on two parameters $R_{t}$ and 
$\sin \beta$ (or $\lambda_{t}^{G}$ and $\tan \beta $).  In general, 
the present model contains four parameters: 
$\epsilon_{G}$, $\epsilon_{P}$, $\tan \beta = v_{2}/v_{1}$, and 
$\lambda_{t}^{G} = 81\lambda_{b}^{G} = 81\lambda_{\tau}^{G} = 
\frac{2}{3}\lambda_{H}$.  It is not difficult to notice that 
$\epsilon_{G}$ and  $\epsilon_{P}$ are  
determined solely by the Clebsch factors and mass ratios of the charged leptons 
\begin{eqnarray}
& & \epsilon_{G} = \left(\frac{v_{5}}{v_{10}}\right) 
\sqrt{\left(\frac{r_{2}}{r_{3}}\right)}  = 
\sqrt{\frac{m_{\mu}}{m_{\tau}} \frac{\eta_{\tau}}{\eta_{\mu}}
\frac{w_{e}}{3y_{e}}} = 2.987 \times 10^{-1}, \\ 
& & \epsilon_{P} = \left(\frac{v_{5}}{\bar{M}_{P}}\right) 
\sqrt{\left(\frac{r_{1}}{r_{3}}\right)} = 
\left( \frac{4}{9} \frac{m_{e}m_{\mu}}{m_{\tau}^{2}}
\frac{\eta_{\tau}^{2}}{\eta_{e}\eta_{\mu}} \frac{w_{e}^{2}}{z_{e}^{2}} 
\right)^{1/4} = 1.011 \times 10^{-2}.
\end{eqnarray}
The coupling $\lambda_{t}^{G}$ (or $R_{t}$) can be determined by the mass ratio 
of the bottom quark and $\tau$ lepton
\begin{eqnarray}
& & \lambda_{t}^{G} = \frac{1}{\sqrt{K_{t}}} \frac{\sqrt{1 -
R_{t}^{-12}}}{R_{t}^{-6}} = 1.25\ \zeta_{t}, \\
& &  \zeta_{t} \equiv \left( \sqrt{\frac{8.65}{K_{t}}}\right) \left(
\frac{0.80}{R_{t}^{-1}}\right)^{6} \left( \frac{\sqrt{1 -
R_{t}^{-12}}}{0.965}\right) , \\
& & R_{t}^{-1} = \frac{m_{b}}{m_{\tau}} \frac{\eta_{\tau}}{\eta_{b}}
\frac{1}{\eta_{D/E}} = 0.80\  \left( 
\frac{m_{b}(m_{b})}{4.25 GeV}\right) \left(
\frac{1.49}{\eta_{b}}\right) \left( \frac{2.04}{\eta_{D/E}} \right).
\end{eqnarray}
$\tan \beta$ is fixed by the $\tau$ lepton mass 
\begin{eqnarray} 
& & \cos \beta = \frac{m_{\tau} \sqrt{2}}{\eta_{E} \eta_{\tau} v 
\lambda_{\tau}^{G}} 
= \left(\frac{0.41}{\zeta_{t}}\right) \left(\frac{3^{n}}{81}\right), 
\nonumber \\
& &  \sin \beta = \sqrt{ 1- (\frac{0.41}{\zeta_{t}}\frac{3^{n}}{81})^{2}} = 
0.912\  \left( \frac{\sqrt{ 1- (\frac{0.41}{\zeta_{t}}
\frac{3^{n}}{81})^{2}}}{0.912}\right), 
\nonumber \\
& & \tan \beta = 2.225\  \left(\frac{81}{3^{n}}\right) 
\left( \frac{\sqrt{\zeta_{t}^{2} -
(0.41)^{2}(3^{n}/81)^{2}}}{0.912}\right) .
\end{eqnarray}
With these considerations, the top quark mass is given by
\begin{equation}
m_{t}(m_{t}) = 173.4\  \left( \frac{\eta_{U}}{3.33}\right) 
\left( \sqrt{\frac{8.65}{K_{t}}}\right)  
\left(\frac{\sqrt{1-R_{t}^{-12}}}{0.965} \right) \left( \frac{\sqrt{1 -
(0.41/\zeta_{t})^{2}}}{0.912}\right) \  GeV   
\end{equation}

 Given  $\epsilon_{G}$ and $\epsilon_{P}$ as
well  $\lambda_{t}^{G}$, the Yukawa coupling 
matrices of the fermions at the GUT scale are then
known. It is of interest to expand the above fermion Yukawa 
coupling matrices $\Gamma_{f}^{G}$ in terms of the 
parameter $\lambda=0.22$ (the Cabbibo angle), 
as  Wolfenstein \cite{WOLFENSTEIN} did 
 for  the CKM mixing matrix. 
\begin{eqnarray}
& & \Gamma_{u}^{G} = 1.25\zeta_{t}  \left( \begin{array}{ccc}
0  & 0.60\lambda^{6}  & 0  \\
1.35\lambda^{6}  & 0  & -0.89\lambda^{2} \\
0  &  -0.89\lambda^{2}  &  1 
\end{array}  \right), \\
& &  \Gamma_{d}^{G} = -\frac{1.25\zeta_{t}}{81}
\left( \begin{array}{ccc}
0  &  1.77\lambda^{4}  & 0  \\
1.77\lambda^{4}  & 0.41 \lambda^{2}e^{i\frac{\pi}{2}}  & -1.09\lambda^{3} \\
0  &  -1.09\lambda^{3}  &  1 
\end{array}  \right), \\
& & \Gamma_{e}^{G} = -\frac{1.25\zeta_{t}}{81} \left( \begin{array}{ccc}
0  & 1.77\lambda^{4}  & 0  \\
1.77\lambda^{4}  & 1.23 \lambda^{2}e^{i\frac{\pi}{2}}  & 1.40\lambda^{2} \\
0  &  1.40\lambda^{2}  &  1 
\end{array}  \right), \\  
& & \Gamma_{\nu}^{G} = -\left(\frac{1.25\zeta_{t}}{81}\right) \left(
\frac{2.581}{5^{5}}\right)  \left( \begin{array}{ccc}
0  & 1 & 0  \\
1  & 0.86 \lambda^{3}e^{i\frac{\pi}{2}}  & 1.472\lambda^{4} \\
0  &  1.472\lambda^{4}  &  1.757 \lambda
\end{array}  \right) 
\end{eqnarray}   

  Using the CKM parameters and quark masses  predicted in the present 
model,  the bag parameter $B_{K}$ can be  
extracted from the indirect CP-violating 
parameter $|\varepsilon_{K}|=2.6 \times 10^{-3}$ 
in $K^{0}$-$\bar{K}^{0}$ system via
\begin{equation}
B_{K} = 0.90\ \left(\frac{0.57}{\eta_{2}}\right) \left(
\frac{|\varepsilon_{K}|}{2.6\times 10^{-3}}\right) \left(\frac{0.138
y_{t}^{1.55}}{A^{4}(1-\rho)\eta }\right) \left( \frac{1.41}{1 + \frac{0.246
y_{t}^{1.34}}{A^{2}(1-\rho)}}\right)
\end{equation}
The B-meson decay constant can also be obtained from fitting 
the $B^{0}$-$\bar{B}^{0}$ mixing 
\begin{equation}
f_{B}\sqrt{B} = 207\ \left(\sqrt{\frac{0.55}{\eta_{B}}}\right) \left(
\frac{\Delta M_{B_{d}}(ps^{-1})}{0.465}\right) \left(\frac{0.77
y_{t}^{0.76}}{A \sqrt{(1-\rho)^{2} + \eta^{2}}}\right)\ MeV 
\end{equation}  
with $y_{t} = 175 GeV /m_{t}(m_{t})$ and $\eta_{2}$ and $\eta_{B}$ being the
QCD corrections\cite{ETA}. Note that we did not consider the possible
contributions to $\varepsilon_{K}$ and $\Delta M_{B_{d}}$ from box diagrams 
through exchanges of superparticles. To have a complete analysis, these 
contributions should be included in a more detailed consideration in the 
future. The parameter $B_{K}$ was estimated ranging from $1/3$ to 1 based on
various approaches. Recent analysis using the lattice 
methods \cite{LATTICE,BK} gives 
$B_{K} = 0.82 \pm 0.1$ .  There are also various calculations on the parameter
$f_{B_{d}}$. From the recent lattice analyses \cite{LATTICE,FB}, 
$f_{B_{d}} = (200\pm 40)$ MeV, $B_{B_{d}} = 1.0 \pm 0.2$.
 QCD sum rule calculations\cite{QCDSR} also gave a compatible result. 
An interesting upper bound \cite{YLWU1} 
$f_{B}\sqrt{B} < 213 $MeV for
$m_{c} = 1.4$GeV and $m_{b} = 4.6$ GeV or $f_{B}\sqrt{B} < 263 $MeV for
$m_{c} = 1.5$GeV and $m_{b} = 5.0$ GeV has been obtained by relating the
hadronic mixing matrix element, $\Gamma_{12}$, to the decay rate of the bottom
quark. 

  The direct CP-violating parameter Re($\varepsilon'/\varepsilon)$ 
in the K-system has been estimated by the standard method. 
The uncertanties mainly arise from the 
hadronic matrix elements\cite{PASCHOS}. 
We have included the next-to-leading  order
contributions from the chiral-loop\cite{CL1,CL2,YLWU2} and 
the next-to-leading order perturbative contributions\cite{NTLM,NTLR} to 
the Wilson coefficients together with  a consistent analysis of 
the $\Delta I =1/2$ rule.  Experimental results on 
Re($\varepsilon'/\varepsilon)$ is inconclusive. The NA31
collaboration at CERN reported a value Re($\varepsilon'/\varepsilon)=
(2.3 \pm 0.7)\cdot 10^{-3}$ \cite{NA31} which clearly indicates direct CP
violation, while the value given by E731 at Fermilab, 
Re($\varepsilon'/\varepsilon)= (0.74 \pm 0.59)\cdot 10^{-3}$ \cite{E731} is
compatible with superweak theories \cite{SW} in which 
$\varepsilon'/\varepsilon=0$.  The average value quoted in \cite{PDG} is 
Re($\varepsilon'/\varepsilon)= (1.5 \pm 0.8)\cdot 10^{-3}$\footnote{ Recently, two improved 
experiments have reported new results:  $Re(\varepsilon'/\varepsilon) = (18.5\pm 4.5\pm 5.8)\times 10^{-4}$  
  by the NA48 collaboration at CERN \cite{NA48}, and 
  $Re(\varepsilon'/\varepsilon) = (28.0\pm 3.0\pm 2.8)\times 10^{-4}$ 
  by the KTeV collaboration at Fermilab\cite{KTEV}, where only $23 \%$ data have been analyzed. 
The hadronic matrix elements have been reanalyzed with paying attention to the matching between 
QCD and Chiral Perturbation Theory (ChPT), as a consequence, both direct CP-violating parameter
 $ \varepsilon'/\varepsilon$ and $\Delta I = 1/2$ rule can be consistently predicted\cite{YLW}. 
With the improved hadronic matrix elements, the present predicted values for the mixing angles and 
CP-violating phase lead to a bigger value for the ratio $ \varepsilon'/\varepsilon$, numerically, 
it is found to be $ \varepsilon'/\varepsilon \simeq 28\times 10^{-4}$.  }
 
  For predicting physical observables, it is better to use $J_{CP}$ , 
the rephase-invariant CP-violating quantity, together with 
$\alpha$, $\beta$ and $\gamma$ , the three angles of
the unitarity triangle  of a three-family CKM matrix
\begin{equation}
\alpha = arg. \left( -\frac{V_{td}V_{tb}^{\ast}}{V_{ud}V_{ub}^{\ast}} \right), 
\quad  \beta = arg. \left( -\frac{V_{cd}V_{cb}^{\ast}}{V_{td}
V_{tb}^{\ast}} \right), \quad 
\gamma = arg. \left( -\frac{V_{ud}V_{ub}^{\ast}}{V_{cd}V_{cb}^{\ast}} \right)
\end{equation}
where $\sin 2\alpha $, $\sin 2\beta$ and $\sin 2 \gamma$ can in principle be
measured in $B^{0}/\bar{B}^{0} \rightarrow \pi^{+} \pi^{-}$ \cite{ALPHA}, 
$J/\psi K_{S}$\cite{BETA} and $B^{-} \rightarrow K^{-} D$\cite{GAMMA}, 
respectively.  $|V_{us}|$ has been extracted with good accuracy from 
$K\rightarrow \pi e \nu $ and
hyperon decays \cite{PDG}. $|V_{cb}|$  can be determined from both exclusive 
and inclusive semileptonic $B$ decays with  values given by  
\begin{equation}
|V_{cb}| = \left\{ \begin{array}{lll}
0.0406 \pm 0.00036\ (exp.) \pm 0.0016\ (theor.);  & \mbox{ ICHEP2000\cite{ICHEP} },  \\
0.0389 \pm 0.0005_{exp} \pm 0.0020_{theor.}; & \mbox{ HQEFT\cite{YWW,WY}} 
\end{array} \right.
\end{equation}
from inclusive semileptonic $B$ decays and 
\begin{equation}
|V_{cb}| = \left\{ \begin{array}{llll}
  0.0418\pm 0.0016\ (exp.) \pm 0.0021\ (theor.); & \mbox{\cite{ICHEP} }\\
0.0404\pm 0.0015\ (exp.) \pm 0.0020\ (theor.); \\
0.0395\pm 0.0029\ (exp.) \pm 0.0019\ (theor.); & \mbox{HQEFT\cite{WWY} } \\
0.0382\pm 0.0041\ (exp.) \pm 0.0028\ (theor.); &\mbox{HQEFT \cite{WWY} } 
\end{array}
\right.
\end{equation}
from exclusive semileptonic $B$ decays. Where the first result was be obtained for 
$F(1) =0.88 \pm 0.04$ and the second for  $F(1) =0.91 \pm 0.04$. The third one 
is from $B\rightarrow D^{\ast} l \nu$ and the last one from  $B\rightarrow D l \nu$.

 Another CKM parameter $|V_{ub}/V_{cb}|$ is extracted from a study of the
semileptonic $B$ decays near the end point region of the lepton spectrum. 
The present experimental measurements are compatible with\cite{ICHEP,YWW}
\begin{equation}
|\frac{V_{ub}}{V_{cb}}| = 0.08 \pm 0.01\ (exp.) \pm 0.02\ (theor.)
\end{equation}
The CKM parameter $|V_{td}/V_{ts}|$ is constrained by the indirect
CP-violating parameter $|\varepsilon | $ in kaon decays and
$B^{0}$-$\bar{B}^{0}$ mixing $x_{d}$. Large uncertainties of $|V_{td}/V_{ts}|$ 
are caused by the bag parameter $B_{K}$ and the leptonic $B$ decay 
constant $f_{B}$.

{\bf  TABLE II.}  Output observables and model parameters and their 
predicted values with input parameters  $m_{e}$ = 0.511 MeV, 
$m_{\mu}$ = 105.66 MeV, $m_{\tau}$ = 1.777 GeV and $m_{b}(m_{b})$ = 4.25 GeV.
\\
{\small
\begin{tabular}{|c|c|c|c|c|}   \hline
 Output parameters   &  Output values   &  Data\cite{TOP,MASS,PDG} & 
 Output para.   &  Output values    \\ \hline 
$M_{t}$\ [GeV]  &  182   &  $175\pm 6 $  &  $J_{CP} = A^{2} 
\lambda^{6} \eta $ & $2.68 \times 10^{-5}$  \\
$m_{c}(m_{c})$\ [GeV]  &  1.27   & $1.27 \pm 0.05$  & 
 $\alpha$ & $86.28^{\circ}$ \\ 
$m_{u}$(1GeV)\ [MeV]  &  4.31   &  $4.75 \pm 1.65$ & 
$\beta$ & $22.11^{\circ}$ \\
$m_{s}$(1GeV)\ [MeV]  &  156.5  &  $165\pm 65$  &  
$\gamma$ & $71.61^{\circ}$  \\
$m_{d}$(1GeV) \ [MeV]  &  6.26 & $8.5 \pm 3.0$ & 
$m_{\nu_{\tau}}$ [eV]  & $ 2.4504$    \\
$|V_{us}|=\lambda $ & 0.22 & $0.221 \pm 0.003$ & 
$m_{\nu_{\mu}}$ [eV]  & $2.4498$   \\
$\frac{|V_{ub}|}{|V_{cb}|} = \lambda \sqrt{\rho^{2} + \eta^{2}}$ & 0.083 & 
$0.08 \pm 0.03$ & $m_{\nu_{e}}$ [eV] & $ 1.27\times 10^{-3}$   \\
$\frac{|V_{td}|}{|V_{ts}|} = \lambda \sqrt{(1-\rho)^{2} + \eta^{2}}$ & 0.209 & 
$0.24 \pm 0.11$ & $m_{\nu_{s}}$ [eV]  & $ 2.8 \times 10^{-3}$  \\
 $|V_{cb}|=A\lambda^{2}$ & 0.0393  &  $0.039 \pm 0.005 $ &
  $|V_{\nu_{\mu}e}| $ &  -0.049  \\
$\lambda_{t}^{G}$  & 1.30  & - &  $|V_{\nu_{e}\tau}| $ &  0.000  \\
$\tan \beta = v_{2}/v_{1}$ & 2.33 & - &   $|V_{\nu_{\tau}e}| $ & -0.049   \\
$\epsilon_{G}$ &  $0.2987$ & - & 
$|V_{\nu_{\mu}\tau}| $ &  -0.707  \\
$\epsilon_{P}$  & $0.0101 $ & - & 
$|V_{\nu_{e}s}|$ & $ 0.038 $ \\
$B_{K}$ & 0.90 &  $0.82 \pm 0.10$ & $M_{N_{1}}$ [GeV] & $\sim 333$  \\
$f_{B}\sqrt{B}$ [MeV] & 207  & $200 \pm 70 $ &
$M_{N_{2}}$ [GeV] & $1.63\times 10^{6}$  \\
 Re($\varepsilon'/\varepsilon)/10^{-3} $ & $1.4 \pm 1.0 $ &  
$1.5 \pm 0.8 $ & $M_{N_{3}}$ [GeV] & $ 333$  
  \\  \hline
\end{tabular}
}
\\

\newpage
 
{\bf  TABLE III.}  Output observables and model parameters and their 
predicted values with input parameters  $m_{e}$ = 0.511 MeV, 
$m_{\mu}$ = 105.66 MeV, $m_{\tau}$ = 1.777 GeV and $m_{b}(m_{b})$ 
= 4.32 GeV.
\\
{\small
\begin{tabular}{|c|c|c|c|c|}   \hline
 Output parameters   &  Output values   &  Data\cite{TOP,MASS,PDG} & 
 Output para.   &  Output values    \\ \hline 
$M_{t}$\ [GeV]  &  179   &  $175\pm 6 $  &  $J_{CP} = A^{2} 
\lambda^{6} \eta $ & $2.62 \times 10^{-5}$  \\
$m_{c}(m_{c})$\ [GeV]  &  1.21   & $1.27 \pm 0.05$  & 
 $\alpha$ & $86.28^{\circ}$ \\ 
$m_{u}$(1GeV)\ [MeV]  &  4.11   &  $4.75 \pm 1.65$ & 
$\beta$ & $22.11^{\circ}$ \\
$m_{s}$(1GeV)\ [MeV]  &  156.5  &  $165\pm 65$  &  
$\gamma$ & $71.61^{\circ}$  \\
$m_{d}$(1GeV) \ [MeV]  &  6.26 & $8.5 \pm 3.0$ & 
$m_{\nu_{\tau}}$ [eV]  & $ 2.4504$    \\
$|V_{us}|=\lambda $ & 0.22 & $0.221 \pm 0.003$ & 
$m_{\nu_{\mu}}$ [eV]  & $2.4498$   \\
$\frac{|V_{ub}|}{|V_{cb}|} = \lambda \sqrt{\rho^{2} + \eta^{2}}$ & 0.083 & 
$0.08 \pm 0.03$ & $m_{\nu_{e}}$ [eV] & $ 1.27\times 10^{-3}$   \\
$\frac{|V_{td}|}{|V_{ts}|} = \lambda \sqrt{(1-\rho)^{2} + \eta^{2}}$ & 0.209 & 
$0.24 \pm 0.11$ & $m_{\nu_{s}}$ [eV]  & $ 2.8 \times 10^{-3}$  \\
 $|V_{cb}|=A\lambda^{2}$ & 0.0389  &  $0.039 \pm 0.005 $ &
  $|V_{\nu_{\mu}e}| $ &  -0.049  \\
$\lambda_{t}^{G}$  & 1.20  & - &  $|V_{\nu_{e}\tau}| $ &  0.000  \\
$\tan \beta = v_{2}/v_{1}$ & 2.12 & - &   $|V_{\nu_{\tau}e}| $ & -0.049   \\
$\epsilon_{G}$ &  $0.2987$ & - & 
$|V_{\nu_{\mu}\tau}| $ &  -0.707  \\
$\epsilon_{P}$  & $0.0101 $ & - & 
$|V_{\nu_{e}s}|$ & $ 0.038 $ \\
$B_{K}$ & 0.96 &  $0.82 \pm 0.10$ & $M_{N_{1}}$ [GeV] & $\sim 361$  \\
$f_{B}\sqrt{B}$ [MeV] & 212  & $200 \pm 70 $ &
$M_{N_{2}}$ [GeV] & $1.77\times 10^{6}$  \\
 Re($\varepsilon'/\varepsilon)/10^{-3} $ & $1.4 \pm 1.0 $ &  
$1.5 \pm 0.8 $ & $M_{N_{3}}$ [GeV] & $ 361$  
  \\  \hline
\end{tabular}
}
\\

{\bf TABLE IV.}  Values of $\eta_{i}$ and $\eta_{F}$ as a 
function of the strong coupling $\alpha_{s}(M_{Z})$
\\

\begin{tabular}{|c|c|c|c|c|c|}  \hline  
$\alpha_{s}(M_{Z})$   &  0.110 & 0.113 & 0.115 & 0.117 & 0.120   \\   \hline 
$\eta_{u,d,s}$  &  2.08 & 2.20  & 2.26 & 2.36 & 2.50  \\
$\eta_{c} $  & 1.90 & 2.00  & 2.05  & 2.12  &  2.25  \\
$\eta_{b}$  & 1.46  &  1.49 & 1.50 & 1.52  & 1.55  \\
$\eta_{e, \mu, \tau}$ & 1.02 & 1.02 & 1.02 & 1.02 & 1.02  \\
$ \eta_{U} $  & 3.26 & 3.33 & 3.38 &  3.44 & 3.50  \\
$ \eta_{D}/\eta_{E}\equiv \eta_{D/E} $ 
& 2.01  &  2.06  &  2.09 &  2.12  & 2.16  \\
$\eta_{E} $  &  1.58 &  1.58 & 1.58 & 1.58 & 1.58   \\  
$\eta_{N}$  & 1.41 & 1.41 & 1.41 & 1.41 & 1.41  \\   \hline
\end{tabular}
\\

    A detail analysis of neutrino masses and mixings will be 
presented in the next section. Before proceeding further, we would like to 
address the following points: Firstly,  given $\alpha_{s}(M_{Z})$ and 
$m_{b}(m_{b})$,  the value of $\tan \beta$ 
depends, as one sees from eq.(56), on the choice of the integer `n' 
in an over all factor $1/3^{n}$, so do the masses of all 
the up-type quarks (see eqs. (48)-(50)). 
For $n > 4$, the value of $\tan \beta $ becomes too small, as a consequence, 
the resulting top quark mass will be below  the present experimental lower 
bound, so do the masses of the up and charm quarks.  In contrast, 
for $1< n < 4$, the values of $\tan \beta$ will become larger, the resulting 
charm quark mass will be above the present upper bound and the top quark
mass is very close to the present upper bound.   Secondly, given 
$m_{b}(m_{b})$ and integer `n', all other quark masses increase 
with $\alpha_{s}(M_{Z})$. This is because the RG scaling 
factors $\eta_{i}$ and $R_{t}$ increase with $\alpha_{s}(M_{Z})$. 
When  $\alpha_{s}(M_{Z})$ is larger than 0.117 and n=4, either 
charm quark mass or bottom quark mass will be above the present upper bound.
Finally, the symmetry breaking direction of the adjoint {\bf 45} $A_{z}$ or 
the Clebsch factor $x_{u}$ is strongly restricted by both 
$|V_{ub}|/|V_{cb}|$ and charm quark mass $m_{c}(m_{c})$.  From these
considerations, we conclude that  the best choice of n 
will be 4 for small $\tan\beta $ and the value of $\alpha_{s}$ should  
around $\alpha_{s}(M_{Z}) \simeq 0.113 $ , which can be seen from table 2b.

\section{Neutrino Masses and Mixings}

   Neutrino masses and mixings , if they exist, are very important 
in astrophysics and crucial for model building. Many 
unification theories predict a see-saw type mass\cite{SEESAW}
$m_{\nu_{i}} \sim m_{u_{i}}^{2}/M_{N}$ with $u_{i} =u, c, t$ being 
up-type quarks. For $M_{N} \simeq (10^{-3}\sim 10^{-4}) M_{GUT} 
\simeq 10^{12}-10^{13}$ GeV, one has 
\begin{equation}
m_{\nu_{e}} < 10^{-7} eV, \qquad m_{\nu_{\mu}} \sim 10^{-3} eV, 
\qquad m_{\nu_{\tau}} \sim (3-21) eV
\end{equation}
In this case solar neutrino anomalous could be explained by 
$\nu_{e} \rightarrow \nu_{\mu}$ oscillation, and the mass of 
$\nu_{\tau}$ is in the range 
relevant to hot dark matter.  However,  LSND events and atmospheric 
neutrino deficit can not be explained in this scenario.    

 By choosing Majorana type Yukawa coupling matrix differently, one can 
construct many models of neutrino mass matrix. As we have shown in the 
Model I that by choosing an appropriate texture structure with some 
diagonal zero elements in the 
right-handed Majorana mass matrix,  
one can explain the recent LSND events, atmospheric neutrino deficit and 
hot dark matter, however, the solar neutrino anomalous can only be 
explained by introducing 
a sterile neutrino.  A similar consideration can be applied to the present
model. The following texture structure with zeros is found to be interesting 
for the present model
\begin{equation}
M_{N}^{G} = M_{R} \left( \begin{array}{ccc} 
0  &  0 &   \frac{1}{2}z_{N}\epsilon_{P}^{2} e^{i(\delta_{\nu} + \phi_{3})}   \\
0  &  y_{N} e^{2i\phi_{2}} & 0 \\
\frac{1}{2}z_{N}\epsilon^{2}_{P} 
e^{i(\delta_{\nu} + \phi_{3})}   & 0 &  w_{N}\epsilon_{P}^{4} e^{2i\phi_{3}} 
\end{array} \right)
\end{equation}   
 The corresponding effective operators are given by 
\begin{eqnarray} 
W_{13}^{N} & = &  \lambda_{1}^{N} 16_{1} (\frac{A_{z}}{v_{5}}) 
(\frac{\bar{\Phi}}{v_{10}})(\frac{\bar{\Phi}}{v_{10}}) 
(\frac{A_{u}}{v_{5}}) 16_{3}\  e^{i(\delta_{\nu} + \phi_{3})} \nonumber  \\
W_{22}^{N} & = & \lambda_{2}^{N}16_{2}(\frac{A_{z}}{A_{X}}) 
(\frac{\bar{\Phi}}{v_{10}})
(\frac{\bar{\Phi}}{v_{10}}) (\frac{A_{z}}{A_{X}}) 
16_{2}\  e^{2i \phi_{2}}   \nonumber  \\ 
W_{33}^{N} & = &  \lambda_{3}^{N}16_{3}(\frac{A_{u}}{v_{5}})
 (\frac{\bar{\Phi}}{v_{10}})
(\frac{\bar{\Phi}}{v_{10}}) (\frac{A_{u}}{v_{5}}) 
16_{3}\  e^{2i\phi_{3}}  \nonumber 
\end{eqnarray}
with $M_{R} = \lambda_{H}\epsilon_{P}^{4} 
\epsilon_{G}^{2}v_{10}^{2}/\bar{M}_{P}$, $\lambda_{1}^{N} =
\epsilon_{P}^{2}M_{R}$, $\lambda_{2}^{N} = M_{R}/\epsilon_{G}^{2}$ and 
$\lambda_{3}^{N} = \epsilon_{P}^{4}M_{R}$.  
It is not difficult to read off the Clebsch factors  
\begin{equation}
y_{N} = 9/25, \qquad  z_{N}= 4, \qquad w_{N} = 16/9 
\end{equation}
where $\delta_{\nu}$, $\phi_{2}$ and $\phi_{3}$ are three phases.
For convenience , we first redefine the phases 
of the three right-handed neutrinos 
$\nu_{R1} \rightarrow e^{i\delta_{\nu}}\nu_{R1}$, 
$\nu_{R2} \rightarrow e^{i\phi_{2}}\nu_{R2}$, and 
$\nu_{R3} \rightarrow e^{i\phi_{3}}\nu_{R3}$,  so that the matrix $M_{N}^{G}$  
becomes real.  

The light neutrino mass matrix is then given 
via see-saw mechanism as follows   
\begin{eqnarray}
M_{\nu} & = & \Gamma_{\nu}^{G} (M_{N}^{G})^{-1} 
(\Gamma_{\nu}^{G})^{\dagger} v_{2}^{2}/2 R_{t}^{-6} \eta_{N}^{2} \nonumber \\ 
& = & M_{0} \left( \begin{array}{ccc} 
1.027 \lambda^{5}  &  -0.88 \lambda^{8} e^{i\frac{\pi}{2}} &  
1.51 \lambda^{9}  \\
-0.88 \lambda^{8} e^{-i\frac{\pi}{2}} &    0.37\lambda^{4} \cos \delta_{\nu}
- 0.456 \lambda^{5} & e^{i\delta_{\nu}}  \\
1.51 \lambda^{9}  & e^{-i\delta_{\nu}}     &  0.49 \lambda^{12}  
\end{array} \right) 
\end{eqnarray}   
with 
\begin{eqnarray}
M_{0} & = & \left(\frac{2}{15^{5}}\right)^{2}
\left(\frac{15}{\epsilon_{P}^{5}}\right)  
\left(\frac{-w_{\nu}z_{\nu}}{y_{N}z_{N}}\right)
\left(\frac{v_{2}^{2}}{2v_{5}}\right)/ R_{t}^{-6} \eta_{N}^{2} \lambda_{H} 
\nonumber \\
& = & 2.45\  \left(\frac{2.36 \times 10^{16} GeV}{v_{5}}\right)
\left(\frac{\zeta_{t}}{1.04}\right)\  eV 
\end{eqnarray} 
It is seen that only one phase, $\delta_{\nu}$ , is  physical. We 
shall assume again  maximum CP violation with $\delta_{\nu} = \pi/2 $ .
Neglecting the small terms of order above $O(\lambda^{7})$, the neutrino 
mass matrix can be simply diagonalized by 
\begin{equation}
 V_{\nu} = \left( \begin{array}{ccc} 
1  &  0 &   0   \\
0 &  c_{\nu}  & -s_{\nu}   \\
0  &  s_{\nu} &  c_{\nu}
\end{array} \right) \left( \begin{array}{ccc} 
1  &  0  &   0   \\
0  &  1  & 0  \\
0  &  0 &  e^{i\delta_{\nu}}
\end{array} \right)
\end{equation}
and the charged lepton mass matrix  by
\begin{equation}   
 V_e = \left( \begin{array}{ccc} 
\bar{c}_1  &  -\bar{s}_1 &   0   \\
\bar{s}_1  &  \bar{c}_1  & 0   \\
0  &  0 &  1
\end{array} \right) \left( \begin{array}{ccc} 
i  &  0  &   0   \\
0  &  c_e  & -s_e   \\
0  &  s_e &  c_e
\end{array} \right)
\end{equation}   
The CKM-type lepton mixing matrix is then given by
\begin{eqnarray}
& & V_{LEP} =  V_{\nu} V_{e}^{\dagger}  = \left(  \begin{array}{ccc} 
V_{\nu_{e}e} & V_{\nu_{e}\mu} & V_{\nu_{e}\tau} \\
V_{\nu_{\mu}e } & V_{\nu_{\mu}\mu} & V_{\nu_{\mu}\tau} \\
V_{\nu_{\tau}e} & V_{\nu_{\tau}\mu} &  V_{\nu_{\tau}\tau} \\
\end{array} \right)  \\ 
& & =  \left(  \begin{array}{ccc} 
\bar{c}_1  &  \bar{s}_1 &   0   \\
-\bar{s}_1 (c_{\nu}c_{e} + s_{\nu}s_{e} e^{i\delta_{\nu}})  &  
\bar{c}_1  (c_{\nu}c_{e} + s_{\nu}s_{e} e^{i\delta_{\nu}})   & 
 -(s_{\nu}c_{e} - c_{\nu}s_{e} e^{i\delta_{\nu}})   \\
-\bar{s}_{1} (s_{\nu}c_{e} - c_{\nu}s_{e} e^{i\delta_{\nu}}) &  
\bar{c}_{1} (s_{\nu}c_{e} - c_{\nu}s_{e} e^{i\delta_{\nu}}) &   
c_{\nu}c_{e} + s_{\nu}s_{e} e^{i\delta_{\nu}}
\end{array} \right)  \nonumber
\end{eqnarray}
where the angles are found to be 
\begin{eqnarray}
& & \tan \bar{\theta}_{1} = \sqrt{\frac{m_{e}}{m_{\mu}}} = 0.0695  \\
& & \tan \theta_{e} = -\frac{x_{e}}{2w_{e}} \epsilon_{G}^{2} = 
-\frac{m_{\mu}}{m_{\tau}} \frac{x_{e}}{6y_{e}} = 0.0149  \\
& & \tan \theta_{\nu} = 1 
\end{eqnarray}
It is of interest to note that these predictions are solely determined without 
involving any new parameters. 

 For masses of light Majorana neutrinos we have
\begin{eqnarray}
m_{\nu_{e}} & = & -\frac{1}{4} 
\frac{z_{\nu}}{w_{\nu}} z_{N} M_{0} = 1.27 \times 10^{-3}\ eV ,  \\
m_{\nu_{\mu}} & = & \left(1 + \frac{15}{2} \frac{z_{\nu}w_{N}}{w_{\nu}z_{N}}
 \epsilon_{P}^{4}\right) M_{0} \simeq 2.4498\ eV \\
m_{\nu_{\tau}} & = & \left(1 - \frac{15}{2} 
\frac{z_{\nu}w_{N}}{w_{\nu}z_{N}} \epsilon_{P}^{4}\right) 
M_{0} \simeq 2.4504\ eV
\end{eqnarray}
The three heavy Majorana neutrinos have masses 
\begin{eqnarray}
& & M_{N_{1}} \simeq M_{N_{3}} \simeq  \frac{1}{2} y_{N} z_{N} 
\epsilon_{P}^{7} v_{5} \lambda_{H}  
\simeq 333 \  \left( \frac{v_{5}}{2.36 \times 10^{16} GeV} \right)\ GeV \\
& & M_{N_{2}}  =  y_{N} \epsilon_{P}^{5} v_{5} \lambda_{H} =  
1.63 \times 10^{6} \  \left( \frac{v_{5}}{2.36 \times 10^{16} GeV} \right)\ GeV
\end{eqnarray}
The RG effects above the GUT scale may be absorbed into the mass 
$M_{0}$. The three heavy Majorana neutrinos in the
present model have their masses  much below  the GUT scale, 
unlike many other GUT models with corresponding masses near  
the GUT scale. In fact,  two of them have masses in the range
comparable with the electroweak scale. 

  As the masses of the three  light neutrinos are very small, a direct
measurement for their masses would be too difficult. An efficient detection 
on light neutrino masses can be achieved through their oscillations. 
The probability that an initial
$\nu_{\alpha}$ of energy $E$ (in unit MeV) gets converted to 
a $\nu_{\beta}$ after travelling a distance $L$ (in unit $m$) is 
\begin{equation}
P_{\nu_{\alpha}\nu_{\beta}} = \delta_{\alpha\beta} - 4 \sum_{j>i} V_{\alpha i}
V_{\beta i}^{\ast} V_{\beta j} V_{\alpha j}^{\ast} \sin^{2} (\frac{1.27 L 
\Delta m_{ij}^{2}}{E}) 
\end{equation}
with $\Delta m_{ij}^{2} = m_{j}^{2} - m_{i}^{2}$ (in unit $eV^{2}$). 
From the above results, we observe the following

1. a $\nu_{\mu} (\bar{\nu}_{\mu}) \rightarrow \nu_{\tau} (\bar{\nu}_{\tau})$
long-wave length oscillation with 
\begin{equation}
\Delta m_{\mu\tau}^{2} = m_{\nu_{\tau}}^{2} - m_{\nu_{\mu}}^{2} \simeq 
2.9 \times 10^{-3} eV^{2}\ , \qquad
\sin^{2}2\theta_{\mu\tau} \simeq 0.987 \ ,
\end{equation}
which could explain the atmospheric neutrino 
with the best fit\cite{SUPERK1}
\begin{equation}
\Delta m_{\mu\tau}^{2} = m_{\nu_{\tau}}^{2} - m_{\nu_{\mu}}^{2} \simeq 
3.0\times  10^{-3} eV^{2}\ , \qquad
\sin^{2}2\theta_{\mu\tau} \simeq 1.0 \ ;
\end{equation}

2. a $\nu_{\mu}(\bar{\nu}_{\mu}) \rightarrow \nu_{e} (\bar{\nu_{e}})$ 
short wave-length oscillation with 
\begin{equation}
\Delta m_{e\mu}^{2} = m_{\nu_{\mu}}^{2} - m_{\nu_{e}}^{2} 
\simeq 6\  eV^{2}, \qquad
\sin^{2}2\theta_{e\mu} \simeq 1.0 \times 10^{-2} \ , 
\end{equation}
could explain the LSND experiment\cite{LSND} 

3. Two massive neutrinos $\nu_{\mu}$ and $\nu_{\tau}$ with 
\begin{equation}
m_{\nu_{\mu}} \simeq m_{\nu_{\tau}}  \simeq 2.45\  eV \ ,
\end{equation}
 fall in the range required by  
possible hot dark matter\cite{DARK}.

4. $(\nu_{\mu} - \nu_{\tau})$ oscillation will be beyond the reach
of CHORUS/NOMAD and E803. However, $(\nu_{e} - \nu_{\tau})$ oscillation
may become interesting as a short wave-length oscillation with 
\begin{equation}
\Delta m_{e\tau}^{2} = m_{\nu_{\tau}}^{2} - m_{\nu_{e}}^{2} 
\simeq 6\  eV^{2}, \qquad
\sin^{2}2\theta_{e\tau} \simeq 1.0 \times 10^{-2} \ , 
\end{equation}
which should provide an independent test on the pattern of the present 
Majorana neutrino mass matrix. 
 
5. Majorana neutrino  allows neutrinoless double beta decay
$(\beta \beta_{0\nu})$\cite{DBETA}. Its decay amplitude is known to
depend on the masses of
Majorana neutrinos $m_{\nu_{i}}$ and the lepton mixing 
matrix elements $V_{ei}$. 
The present model is compatible with the present experimental upper bound
on neutrinoless double beta decay
\begin{equation} 
\bar{m}_{\nu_{e}} = \sum_{i=1}^{3} [ V_{ei}^{2} m_{\nu_{i}} \zeta_{i} ] 
\simeq 1.18 \times 10^{-2}\ eV\ < \  \bar{m}_{\nu}^{upper} \simeq 0.7 \ eV  
\end{equation}  
The decay rate is found to be
\begin{equation}
\Gamma_{\beta\beta} \simeq \frac{Q^{5}G_{F}^{4}\bar{m}_{\nu_{e}}^{2}
p_{F}^{2}}{60\pi^{3}} \simeq 1.0 \times 10^{-61} GeV
\end{equation}
with the two electron energy $Q\simeq 2$ MeV and $p_{F}\simeq 50$ MeV. 

6.  In this case, solar neutrino deficit has to be explained by oscillation 
between $\nu_{e}$ and  a sterile neutrino $\nu_{s}$
\cite{STERILE,CHOUWU,CHOUWU2} via MSW effects. Since 
strong bounds on the number of neutrino species both from the invisible 
$Z^{0}$-width and from primordial nucleosynthesis \cite{NS,NS1} require the 
additional neutrino to be sterile (singlet of SU(2)$\times$ U(1), or 
singlet of SO(10) in the GUT SO(10) model). 
Masses and mixings of the triplet sterile neutrinos can be chosen 
by introducing an additional  singlet scalar with VEV $v_{s}\simeq 336$ GeV.
We find  
\begin{eqnarray}
& & m_{\nu_{s}} = \lambda_{H} v_{s}^{2}/v_{10} \simeq 2.8 \times 10^{-3} eV
\nonumber \\
& & \sin\theta_{es} \simeq \frac{m_{\nu_{L}\nu_{s}}}{m_{\nu_{s}}} 
= \frac{v_{2}}{2v_{s}} \frac{\epsilon_{P}}{\epsilon_{G}^{2}} \simeq 3.8 
\times 10^{-2} 
\end{eqnarray}
with the mixing angle  consistent with the requirement necessary for 
primordial nucleosynthesis \cite{PNS} 
given in \cite{NS}.  The resulting parameters
\begin{equation}
\Delta m_{es}^{2} = m_{\nu_{s}}^{2} - m_{\nu_{e}}^{2} \simeq 6.2 \times 
10^{-6} eV^{2}, \qquad  \sin^{2}2 \theta_{es} \simeq 5.8 \times 10^{-3}
\end{equation}
are consistent with the values \cite{STERILE} obtained from 
fitting the experimental data:
\begin{equation}
\Delta m_{es}^{2} = m_{\nu_{s}}^{2} - m_{\nu_{e}}^{2} \simeq (4-9) \times 
10^{-6} eV^{2}, \qquad  \sin^{2}2 \theta_{es} \simeq (1.6-14) \times 10^{-3}
\end{equation}

\section{Conclusions and Remarks}
 
  Based on the symmetry group SUSY SO(10)$\times \Delta
(48) \times$ U(1)\footnote{Here $\Delta(48)$ is the non-Abelian 
discrete dihedral group $\Delta (3n^2)$ with $n=4$, a subgroup of SU(3)\cite{SUBGROUP}.}  
, we have presented in much greater detail 
an alternative interesting model with small 
$\tan \beta $. The needed nonzero textures are constructed by using the properties of the 
dihedral group $\Delta (48)$ given by Table 5 via Froggatt and Nielsen mechanism \cite{FN} 
shown in Figs.1 and 2. It is amazing that nature has allowed us to 
make predictions in terms of a single Yukawa coupling constant and 
three ratios of the VEVs determined by the structure of the physical 
vacuum and  understand the low energy physics 
from the GUT scale physics. It has also suggested that nature favors 
 maximal spontaneous CP violation. In comparison with the model 
with large $\tan \beta \sim m_{t}/m_{b}$, i.e., Model I,   
the model analyzed here with low $\tan \beta$, i.e., Model II 
has provided a consistent 
picture on the 23 parameters with better accuracy. Besides, 
ten relations involving fermion masses and CKM matrix elements 
are obtained with four of them independent of the RG scaling effects.
Five relations in the light neutrino sector are also found to be independent of
the RG scaling effects. 
These relations are our main results which contain only low energy observables.
As an analogy to the Balmer series formula, these relations may remain to 
be considered as empirical at the present moment. 
They have been tested by the existing
experimental data to a good approximation   and can be tested further directly
by more precise experiments in the future.  The two types of the 
models corresponding to the large $\tan\beta$ (Model I) 
and low $\tan \beta$ (Model II) 
might  be distinguished in testing the MSSM Higgs sector at Colliders as well
as by precisely measuring the ratio $|V_{ub}/V_{cb}|$ since this ratio does
not receive radiative corrections in both models. 
The neutrino sector is of special interest for further study.
Though the recent LSND experiment, atmospheric neutrino deficit,
and hot dark matter could be simultaneously explained
in the present model, solar neutrino puzzle can be 
understood  only by introducing an SO(10) singlet sterile neutrino.  
The scenario for the neutrino sector 
can be further tested through ($\nu_{e}-\nu_{\tau}$) and 
($\nu_{\mu}-\nu_{\tau}$) oscillations since the present scenario has predicted 
a short wave  ($\nu_{e}-\nu_{\tau}$) oscillation. However, the 
($\nu_{\mu}-\nu_{\tau}$) oscillation is beyond the reach of CHORUS/NOMAD and
E803. As we have also shown that one may abondon the assumption of universality 
for all terms in the superpotential and use a permutation symmetry among the
fields concerning the `22' and `32' textures, consequently, the resulting
predictions in the quark and charged lepton sector are unchanged and remain
involving four parameters. It is also interesting to note that even if without
imposing the permutation symmetry and universality of the coupling 
constants, the resulting Yukawa coupling 
matrices of the quarks and leptons only add one additional parameter which can 
be determined by the charm quark mass. For $m_{c}(m_{c}) = 1.27\pm 0.05$GeV,
the resulting predictions remain the same as those in Tables II and III for 
the quark and charged lepton sector. For the neutrino sector, three 
additional parameters corresponding to the three nonzero textures are involved. 
Nevertheless, it remains amazing that nature allows us 
to make predictions on 23 observables by only using nine parameters for this
general case. It is expected that more precise measurements from 
CP violation, neutrino oscillation  and various low energy experiments 
in the near future could provide an important test on the present model and 
guide us to a more fundamental theory. It is of interest to note that with gauged SO(3) 
flavor symmetry, the intriguing bi-maximal mixing scenario can be resulted to understand 
both solar and atmospheric neutrino data\cite{YLWU}.



\setlength{\unitlength}{0.7cm}

\begin{picture}(25,19.7)

\thicklines

\put(1,18.5){\framebox(2.3,1){$(1)$: $``33"$}}
\put(7.08,17){\vector(1,0){1}}
\put(8.08,17){\line(1,0){0.84}}
\multiput(9,17)(0,0.3){9}{\line(0,1){0.25}}
\put(8.85,19.6){${\bf \times}$}
\put(9.08,17){\line(1,0){0.84}}
\put(10.92,17){\vector(-1,0){1}}
\multiput(11,17)(0,0.3){9}{\line(0,1){0.25}}
\put(10.8,19.8){${\bf \chi }$}
\put(11.08,17){\line(1,0){0.84}}
\put(12.92,17){\vector(-1,0){0.85}}
\put(12.07,17){\vector(-1,0){0.15}}
\put(4,16.9){$ T_{1}$}
\put(9.2,19.6){$ \left \langle {10_{1}} \right \rangle $}
\put(10.2,16.5){$T_{2}$}
\put(11.2,19.6){$ \left \langle {\bar{T}_{3}^{(3)}} \right \rangle $}
\put(13.2,16.9){$T_{1}$}
\put(5.08,17){\vector(1,0){0.85}}
\put(5.93,17){\vector(1,0){0.15}}
\put(6.08,17){\line(1,0){0.84}}
\multiput(7,17)(0,0.3){9}{\line(0,1){0.25}}
\put(6.8,19.8){${\bf \chi_{1}}$}
\put(7.2,19.6){$ \left \langle {\bar{T}_{3}} \right \rangle $}
\put(7.5,16.5){$T_{2}$}

\put(1,13.5){\framebox(2.3,1){$(2)$: $``32"$}}
\put(7.08,12){\vector(1,0){1}}
\put(8.08,12){\line(1,0){0.84}}
\multiput(9,12)(0,0.3){9}{\line(0,1){0.25}}
\put(8.85,14.6){${\bf \times}$}
\put(9.08,12){\line(1,0){0.84}}
\put(10.92,12){\vector(-1,0){1}}
\multiput(11,12)(0,0.3){9}{\line(0,1){0.25}}
\put(10.8,14.8){${\bf \chi }$}
\put(11.08,12){\line(1,0){0.84}}
\put(12.92,12){\vector(-1,0){0.85}}
\put(12.07,12){\vector(-1,0){0.15}}
\put(4,11.9){$ T_{1}$}
\put(9.2,14.6){$ \left \langle {10_{1}} \right \rangle $}
\put(10.2,11.5){$T_{3}$}
\put(11.2,14.6){$ \left \langle {\bar{T}_{3}^{(3)}} \right \rangle $}
\put(13.2,11.9){$T_{1}$}
\put(5.08,12){\vector(1,0){0.85}}
\put(5.93,12){\vector(1,0){0.15}}
\put(6.08,12){\line(1,0){0.84}}
\multiput(7,12)(0,0.3){9}{\line(0,1){0.25}}
\put(6.8,14.8){${\bf \chi_{2}}$}
\put(7.2,14.6){$ \left \langle {T_{3}} \right \rangle $}
\put(7.5,11.5){$\bar{T}_{3}$}

\put(1,8.5){\framebox(2.3,1){$(3)$: $``22"$}}
\put(7.08,7){\vector(1,0){1}}
\put(8.08,7){\line(1,0){0.84}}
\multiput(9,7)(0,0.3){9}{\line(0,1){0.25}}
\put(8.85,9.6){${\bf \times}$}
\put(9.08,7){\line(1,0){0.84}}
\put(10.92,7){\vector(-1,0){1}}
\multiput(11,7)(0,0.3){9}{\line(0,1){0.25}}
\put(10.8,9.8){${\bf \chi }$}
\put(11.08,7){\line(1,0){0.84}}
\put(12.92,7){\vector(-1,0){0.85}}
\put(12.07,7){\vector(-1,0){0.15}}
\put(4,6.9){$ T_{1}$}
\put(9.2,9.6){$ \left \langle {10_{1}} \right \rangle $}
\put(10.2,6.5){$T_{3}$}
\put(11.2,9.6){$ \left \langle {\bar{T}_{3}^{(3)}} \right \rangle $}
\put(13.2,6.9){$T_{1}$}
\put(5.08,7){\vector(1,0){0.85}}
\put(5.93,7){\vector(1,0){0.15}}
\put(6.08,7){\line(1,0){0.84}}
\multiput(7,7)(0,0.3){9}{\line(0,1){0.25}}
\put(6.8,9.8){${\bf \chi_{3}}$}
\put(7.2,9.6){$ \left \langle {\bar{T}_{1}} \right \rangle $}
\put(7.5,6.5){$\bar{T}_{3}$}

\put(1,3.5){\framebox(2.3,1){$(0)$: $``12"$}}
\put(7.08,2){\vector(1,0){1}}
\put(8.08,2){\line(1,0){0.84}}
\multiput(9,2)(0,0.3){9}{\line(0,1){0.25}}
\put(8.85,4.6){${\bf \times}$}
\put(9.08,2){\line(1,0){0.84}}
\put(10.92,2){\vector(-1,0){1}}
\multiput(11,2)(0,0.3){9}{\line(0,1){0.25}}
\put(10.8,4.8){${\bf \chi }$}
\put(11.08,2){\line(1,0){0.84}}
\put(12.92,2){\vector(-1,0){0.85}}
\put(12.07,2){\vector(-1,0){0.15}}
\put(4,1.9){$ T_{1}$}
\put(9.2,4.6){$ \left \langle {10_{1}} \right \rangle $}
\put(10.2,1.5){$T_{3}$}
\put(11.2,4.6){$ \left \langle {\bar{T}_{3}^{(3)}} \right \rangle $}
\put(13.2,1.9){$T_{1}$}
\put(5.08,2){\vector(1,0){0.85}}
\put(5.93,2){\vector(1,0){0.15}}
\put(6.08,2){\line(1,0){0.84}}
\multiput(7,2)(0,0.3){9}{\line(0,1){0.25}}
\put(6.8,4.8){${\bf \chi_{0}}$}
\put(7.2,4.6){$ \left \langle {T_{2}} \right \rangle $}
\put(7.5,1.5){$\bar{T}_{3}$}

\end{picture}

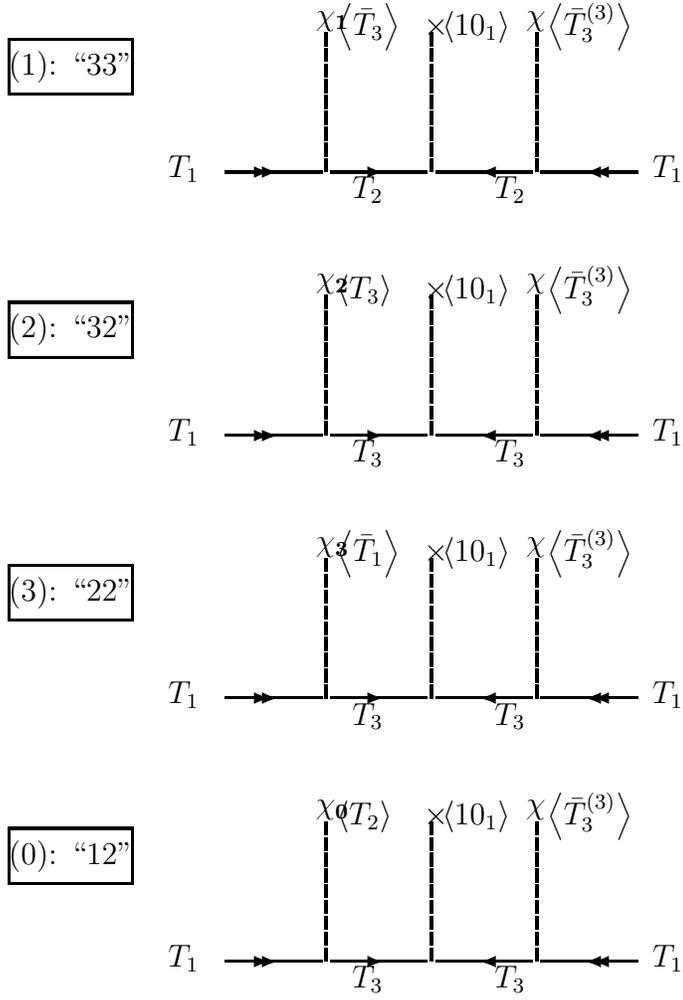
{\bf FIG. 1.} four non-zero textures resulting from the family symmetry $\Delta
(48)$ and U(1) symmetry, are needed for constructing fermion Yukawa 
coupling matrices.


\newpage 



\setlength{\unitlength}{0.7cm}

\begin{picture}(25,19.7)

\thicklines

\put(1,18.5){\framebox(2.3,1){$(1)$: $``33"$}}
\put(7.08,17){\vector(1,0){1}}
\put(8.08,17){\line(1,0){0.84}}
\multiput(9,17)(0,0.3){9}{\line(0,1){0.25}}
\put(8.85,19.6){${\bf \times}$}
\put(9.08,17){\line(1,0){0.84}}
\put(10.92,17){\vector(-1,0){1}}
\multiput(11,17)(0,0.3){9}{\line(0,1){0.25}}
\put(10.8,19.8){${\bf \chi' }$}
\put(11.08,17){\line(1,0){0.84}}
\put(12.92,17){\vector(-1,0){0.85}}
\put(12.07,17){\vector(-1,0){0.15}}
\put(4,16.9){$ T_{1}$}
\put(9.2,19.6){$ \left \langle \bar{\Phi}\bar{\Phi}\right \rangle $}
\put(10.2,16.5){$\bar{T}_{3}$}
\put(11.2,19.6){$ \left \langle {T_{3}^{(2)}} \right \rangle $}
\put(13.2,16.9){$T_{1}$}
\put(5.08,17){\vector(1,0){0.85}}
\put(5.93,17){\vector(1,0){0.15}}
\put(6.08,17){\line(1,0){0.84}}
\multiput(7,17)(0,0.3){9}{\line(0,1){0.25}}
\put(6.8,19.8){${\bf \chi'_{1}}$}
\put(7.2,19.6){$ \left \langle {\bar{T}_{1}} \right \rangle $}
\put(7.5,16.5){$T_{3}$}

\put(1,13.5){\framebox(2.3,1){$(2)$: $``13"$}}
\put(7.08,12){\vector(1,0){1}}
\put(8.08,12){\line(1,0){0.84}}
\multiput(9,12)(0,0.3){9}{\line(0,1){0.25}}
\put(8.85,14.6){${\bf \times}$}
\put(9.08,12){\line(1,0){0.84}}
\put(10.92,12){\vector(-1,0){1}}
\multiput(11,12)(0,0.3){9}{\line(0,1){0.25}}
\put(10.8,14.8){${\bf \chi' }$}
\put(11.08,12){\line(1,0){0.84}}
\put(12.92,12){\vector(-1,0){0.85}}
\put(12.07,12){\vector(-1,0){0.15}}
\put(4,11.9){$ T_{1}$}
\put(9.2,14.6){$ \left \langle \bar{\Phi}\bar{\Phi}\right \rangle $}
\put(10.2,11.5){$\bar{T}_{3}$}
\put(11.2,14.6){$ \left \langle {T}_{3}^{(2)} \right \rangle $}
\put(13.2,11.9){$T_{1}$}
\put(5.08,12){\vector(1,0){0.85}}
\put(5.93,12){\vector(1,0){0.15}}
\put(6.08,12){\line(1,0){0.84}}
\multiput(7,12)(0,0.3){9}{\line(0,1){0.25}}
\put(6.8,14.8){${\bf \chi'_{2}}$}
\put(7.2,14.6){$ \left \langle {T_{2}} \right \rangle $}
\put(7.5,11.5){$T_{3}$}

\put(1,8.5){\framebox(2.3,1){$(3)$: $``22"$}}
\put(7.08,7){\vector(1,0){1}}
\put(8.08,7){\line(1,0){0.84}}
\multiput(9,7)(0,0.3){9}{\line(0,1){0.25}}
\put(8.85,9.6){${\bf \times}$}
\put(9.08,7){\line(1,0){0.84}}
\put(10.92,7){\vector(-1,0){1}}
\multiput(11,7)(0,0.3){9}{\line(0,1){0.25}}
\put(10.8,9.8){${\bf \chi' }$}
\put(11.08,7){\line(1,0){0.84}}
\put(12.92,7){\vector(-1,0){0.85}}
\put(12.07,7){\vector(-1,0){0.15}}
\put(4,6.9){$ T_{1}$}
\put(9.2,9.6){$ \left \langle \bar{\Phi}\bar{\Phi}\right \rangle $}
\put(10.2,6.5){$T_{2}$}
\put(11.2,9.6){$ \left \langle {T}_{3}^{(2)} \right \rangle $}
\put(13.2,6.9){$T_{1}$}
\put(5.08,7){\vector(1,0){0.85}}
\put(5.93,7){\vector(1,0){0.15}}
\put(6.08,7){\line(1,0){0.84}}
\multiput(7,7)(0,0.3){9}{\line(0,1){0.25}}
\put(6.8,9.8){${\bf \chi'_{3}}$}
\put(7.2,9.6){$ \left \langle {T_{3}} \right \rangle $}
\put(7.5,6.5){$T_{2}$}

\end{picture}

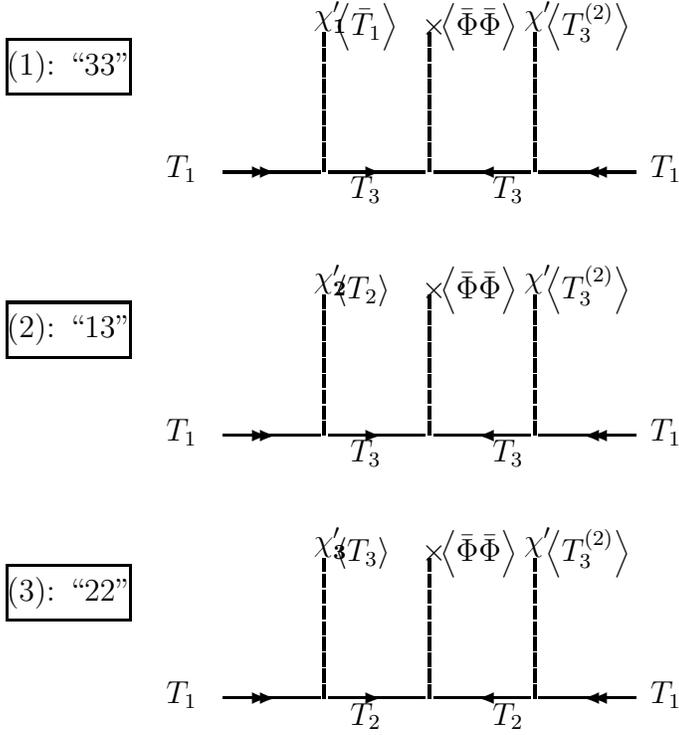
{\bf FIG. 2.} three non-zero textures resulting from the family symmetry $\Delta
(48)$ and U(1) symmetry, are needed for constructing right-handed 
Majorana neutrino mass matrix.


\vspace{1.5cm}
{\bf TABLE V.},  Decomposition of the product of two triplets, 
$T_{i}\otimes T_{j}$ and $T_{i}\otimes \bar{T}_{j}$ in $\Delta (48)$. 
Triplets $T_{i}$ and $\bar{T}_{i}$ are simply denoted by $i$ and $\bar{i}$
respectively. For example $T_{1}\otimes \bar{T}_{1} = A \oplus T_{3} \oplus 
\bar{T}_{3} \equiv A3\bar{3}$, here $A$ represents singlets.
\\

\begin{tabular}{|c|ccccc|}  \hline
$\Delta (48)$ & 1   & $\bar{1}$  &  2  &  3  & $\bar{3}$  \\ \hline
  1        & $\bar{1} \bar{1}$2  & A3$\bar{3}$  & $\bar{1} 3 \bar{3}$ & 
 123  & 
12$\bar{3}$  \\ 
2   &  $\bar{1}3\bar{3}$  & 13$\bar{3}$   &  A22  & 1$\bar{1}\bar{3}$  & 
1$\bar{1}$3 \\
3  &   123  & $\bar{1}$23  & 1$\bar{1}\bar{3}$  & 2$\bar{3}\bar{3}$  & 
A1$\bar{1}$ \\    \hline
\end{tabular}
\\



\end{document}